\documentclass[12pt]{article}
\usepackage{amsmath}
\usepackage{graphicx}
\usepackage{enumerate}
\usepackage{natbib}
\usepackage{url} 

\newcommand{\blind}{1}

\renewcommand{\baselinestretch}{1.5}

\def\boxit#1{\vbox{\hrule\hbox{\vrule\kern6pt
          \vbox{\kern6pt#1\kern6pt}\kern6pt\vrule}\hrule}}

\def\bse{\begin{eqnarray*}}
\def\ese{\end{eqnarray*}}
\def\be{\begin{eqnarray}}
\def\ee{\end{eqnarray}}
\def\bq{\begin{equation}}
\def\eq{\end{equation}}
\def\bse{\begin{eqnarray*}}
\def\ese{\end{eqnarray*}}

\usepackage{amsmath}
\usepackage[margin=1in,headheight=0.45in]{geometry}
\usepackage[pdftex]{epsfig}
\usepackage{amsfonts}
\usepackage{amssymb}
\usepackage{amsthm}
\usepackage{fancyhdr}
\usepackage{subfig}
\usepackage{float}
\usepackage{bbm}
\usepackage[export]{adjustbox}
\usepackage{multirow}
\usepackage{gensymb}
\usepackage{booktabs}
\usepackage[normalem]{ulem}

\include{graphicsx}
\usepackage[colorlinks=true,linkcolor=blue,citecolor=blue]{hyperref}%

\newtheorem{proposition}{Proposition}[section]

\usepackage{setspace}
\newcommand{\Ltwo}{\mathbbm{L}^2}

\usepackage[usenames,dvipsnames]{color}

\begin{document}

\def\spacingset#1{\renewcommand{\baselinestretch}%
{#1}\small\normalsize} \spacingset{1}



\if1\blind
{
  \baselineskip=28pt \vskip 5mm
    \begin{center} {\LARGE{\bf Evaluating proxy influence in assimilated paleoclimate reconstructions -- Testing the exchangeability of two ensembles of spatial processes}}
    \end{center}

    \baselineskip=14pt \vskip 10mm

    \begin{center}\large
    Trevor Harris\footnote{\baselineskip=12pt Department of Statistics, University of Illinois at Urbana-Champaign},
        Bo Li$^1$,
        Nathan J. Steiger\footnote{\baselineskip=12pt Lamont-Doherty Earth Observatory},
        Jason E. Smerdon$^2$,
        Naveen Narisetty$^1$,
        J. Derek Tucker\footnote{\baselineskip=12pt Sandia National Laboratories, Albuquerque, NM},
    \end{center}
    \baselineskip=19pt \vskip 15mm \centerline{\today} \vskip 6mm
} \fi

\if0\blind
{
  \bigskip
  \bigskip
  \bigskip
  \begin{center}
    {\LARGE\bf Evaluating proxy influence in assimilated paleoclimate reconstructions -- Testing the exchangeability of two ensembles of spatial processes}
\end{center}
  \medskip
} \fi

\begin{abstract}
Climate field reconstructions (CFR) attempt to estimate spatiotemporal fields of climate variables in the past using climate proxies such as tree rings, ice cores, and corals. Data Assimilation (DA) methods are a recent and promising new means of deriving CFRs that optimally fuse climate proxies with climate model output. Despite the growing application of DA-based CFRs, little is understood about how much the assimilated proxies change the statistical properties of the climate model data. To address this question, we propose a robust and computationally efficient method, based on functional data depth, to evaluate differences in the distributions of two spatiotemporal processes. We apply our test to study global and regional proxy influence in DA-based CFRs by comparing the background and analysis states, which are treated as two samples of spatiotemporal fields. We find that the analysis states are significantly altered from the climate-model-based background states due to the assimilation of proxies. Moreover, the difference between the analysis and background states increases with the number of proxies, even in regions far beyond proxy collection sites. Our approach allows us to characterize the added value of proxies, indicating where and when the analysis states are distinct from the background states.
\end{abstract}

\noindent%
\vfill


\newpage
\spacingset{1.5} 
\section{Introduction}
\label{sec:intro}

Since their first high-profile application two decades ago \citep{mann1998global-scale}, multi-proxy spatiotemporal climate field reconstructions (CFRs) have become increasingly popular in the climate science community for their ability to reconstruct global climate variability on seasonal and annual timescales over many hundreds of years into the past \citep{jones2009high, smerdon2016reconstructing, christiansen2017challenges}.  The reconstruction of climate is critical because data from instrumental observations are only available for the past 100-150 years.  CFRs therefore provide estimates of past climate variability and extreme events that may not be well represented over the instrumental interval. This helps to better characterize the physical dynamics of the climate system and how climate may change in the future.

The basic approach of CFRs is to statistically relate a collection of climate proxies, such as isotopic information in ice cores, the width of tree rings, or coral isotope data, to observed climate variables like temperature and soil moisture during their periods of overlap \citep{ip04000e}.
Once the relationship between the proxies and the climate variables is established, the proxies are used to estimate climate variability during periods when observations are not available in the past.  CFRs thus depend critically on the imperfect proxy information and the robustness with which their relationship to observed climate variables can be defined.  A central approach to this problem in the past has been through regularized versions of multivariate regression techniques \citep[e.g.][]{lee2008evaluation, jones2009high, tingley2012piecing, smerdon2012climate, guillot2015statistical, smerdon2016reconstructing, li2016comparison, christiansen2017challenges}. More advanced techniques have been emerging, however, all of which are associated with advantages and challenges that require further evaluation and assessment.

A recent CFR innovation are the paleoclimatic Data Assimilation (DA) algorithms, which are a class of reconstruction methods that optimally combine general circulation models (GCMs) with proxy information to create paleoclimate reconstructions \citep{goosse2012role, steiger2014assimilation, hakim2016LMR, steiger2018PHYDA, tardif2019last}.
The primary advantage of DA approaches is their ability to jointly reconstruct multiple atmosphere-ocean variables and to do so in a manner that is physically consistent within the framework of a climate model. An important distinction of the DA methods, relative to the other statistical (inverse) methods, is the use of forward models that map from climate states to the proxies.
An additional advantage is that DA algorithms naturally provide probabilistic, ensemble estimates of past climate. Such ensemble reconstructions first begin with a background ensemble of states from a climate model. These states are then updated through the equations of DA \citep{steiger2014assimilation}, based on the available proxy information and the uncertainties involved, to arrive at an analysis ensemble state estimate. This probabilistic analysis state provides an uncertainty quantification that is critical given the noisy relationship between paleoclimate proxies and climate variables. 


Despite the rapid development of DA-based reconstruction methods, much remains to be characterized about the influence of each of their two components: climate models and paleoclimate proxies. In currently published DA-based CFRs (e.g., \cite{steiger2018PHYDA}), it is hard to quantify how much information the models and the proxies each contribute to the end product. One approach is independent proxy validation of the analysis states \citep{hakim2016LMR}. Another approach is to compare climate time series and climate patterns in the background and analysis with each other and with observations \citep{singh2018insights}. However, more formal statistical approaches are called for to differentiate whether or not the climate model-based background is fundamentally distinct from the analysis. If the background and analysis are not in fact distinct, then this would imply that DA-based CFRs are essentially dominated by the underlying climate model and fail to glean information from the historical proxy data. A lack of proxy influence would, therefore, indicate a need to fundamentally re-evaluate DA methodologies.


In this paper, we quantify the level of proxy influence in the analysis states of a DA product by introducing a robust and computationally efficient method for evaluating the exchangeability of two ensembles of random fields.
The purpose of this study is therefore twofold: to answer an important climatological question by quantifying and assessing the influence of proxies in a new DA based CFR product and to develop a new statistical test for comparing the distributions of two sets of random fields. 
In the following two subsections, we provide background on the methodological development embodied in this paper and the characteristics of the DA-based CFR that we analyze.

\subsection{Previous work in random fields comparisons} \label{previous}
Comparing two spatial processes has been addressed in both the geostatistics and functional data analysis literature. The general strategy in both frameworks is to reduce the dimension of the random process either by a low-rank decomposition or by parameterization and then to develop a test for evaluating differences in the reduced dimension.

The wavelet decomposition has been widely used to reduce a stochastic process to a finite number of wavelet coefficients, then the comparison between two processes can be transformed into the comparison between two sets of wavelet coefficients \citep{briggs, shen, pavlicova}. \citet{snell} and \citet{wang} introduced methods for comparing random fields based on their spatial interpolation root-mean-square error and R$^2$ coefficient. Their methods were later extended by \citet{hering} to include more arbitrary loss functions. Motivated by \citet{lund} that compared two time series, \citet{li1} proposed a parametric method to jointly assess the first two moments between two random fields. 
 
Functional data analysis approaches assume that the spatial random fields are noisy realizations of an underlying continuous function. 
The majority of existing functional approaches have focused on testing the equality of the mean functions arising from two functional data sets \citep{ramsay, zhang2, horvath, staicu2}, although more recently the second order structure of functional data has also been considered \citep{zhang1}. \citet{li2016comparison} extended \citet{zhang1} to evaluate the joint difference in mean and covariance structure as well as in the trend surface between two spatiotemporal random fields. A nice feature of functional data analysis methods, as opposed to geostatistical methods, is that assumptions about distribution and model specification can be relaxed if there are replicate observations in the data.

All the above procedures are nevertheless inadequate for our problem because the proxies can simultaneously affect the mean, covariance, and higher order structures of the reconstructed climate field. The most comprehensive way to identify proxy influence is therefore to compare the distributions of the background and analysis states. The rich ensemble structure of the background and analysis states also allows us to examine more information than differences in the mean and covariance parameters. We take advantage of the ensembles by employing a functional data approach that is both distribution and parameter free.

The problem of comparing the distributions of functions has remained relatively unexplored. \citet{hall} proposed a Cramer-von Mises-like test by constructing an empirical distribution over each of the samples and measuring the $\Ltwo$ distance between the empirical distributions. \citet{benko} introduced a permutation test on the leading coefficients of the common functional principal components (FPCs) and \citet{corain} introduced three omnibus tests for combining pointwise tests on the observations of the functions. Each of these methods depends on a resampling procedure that renders them computationally prohibitive for large ensembles like the DA ensemble output that we consider.

\citet{staicu} proposed a method based on marginal FPCs that does not require resampling, called the Functional Anderon-Darling ($FAD$) test. The $FAD$ test compares the distributions of the marginal FPCs using the two sample Anderson-Darling test and a Bonferroni correction. \citet{banddepth} proposed a rank based band depth test ($BAND$). The $BAND$ test is closely related to the multivariate distribution test based on the Quality Index
\citep{quality} but it replaces the multivariate simplicial depth \citep{simplicial} with the functional band depth \citep{banddepth}. Both of these tests are inadequate for our data because, as we show in the supplement, they are incapable of detecting heterogeneous variance changes across the domain of the generating process.  This causes the $BAND$ test to experience a severe loss of power and for $FAD$ to miss an important trend (Section \ref{global}) in the our dataset.

In this paper, we propose a new non-parametric statistic, based on the concept of data depth, for assessing the equality of distributions between two spatial data sets. Our test falls into the general category of functional data analysis methods for comparing spatial random fields, but is conceptually different from previous efforts in this area. The use of data depth for comparing two multivariate distributions was first explored by \citet{quality} who introduced the Quality Index ($QI$) for comparing two multivariate distributions. The $QI$ essentially measures the mean outlyingness of one sample from a reference sample using data depth. We will extend their ideas to the functional setting and propose a modification that makes our test statistic invariant to the reference distribution. The use of depth, and particularly Integrated Tukey depth \citep{cuevas}, ensures our test is computationally efficient, distribution free, and invariant to location, scale, warping, and other nuisance properties that could influence the testing \citep{nagy}.

\subsection{Reconstruction data}
The DA-based CFR that we analyze comes from the Paleo Hydrodynamics Data Assimilation product (PHYDA), which is a global paleoclimate reconstruction of both temperature and moisture variables \citep{steiger2018PHYDA}. PHYDA incorporates a simulation from the Community Earth System Model (CESM) last millennium ensemble experiment, run over the historical years 850 C.E. to 1850 C.E. \citep{otto2016climate}.

\begin{figure}
    \centering
    \includegraphics[width=0.6\textwidth]{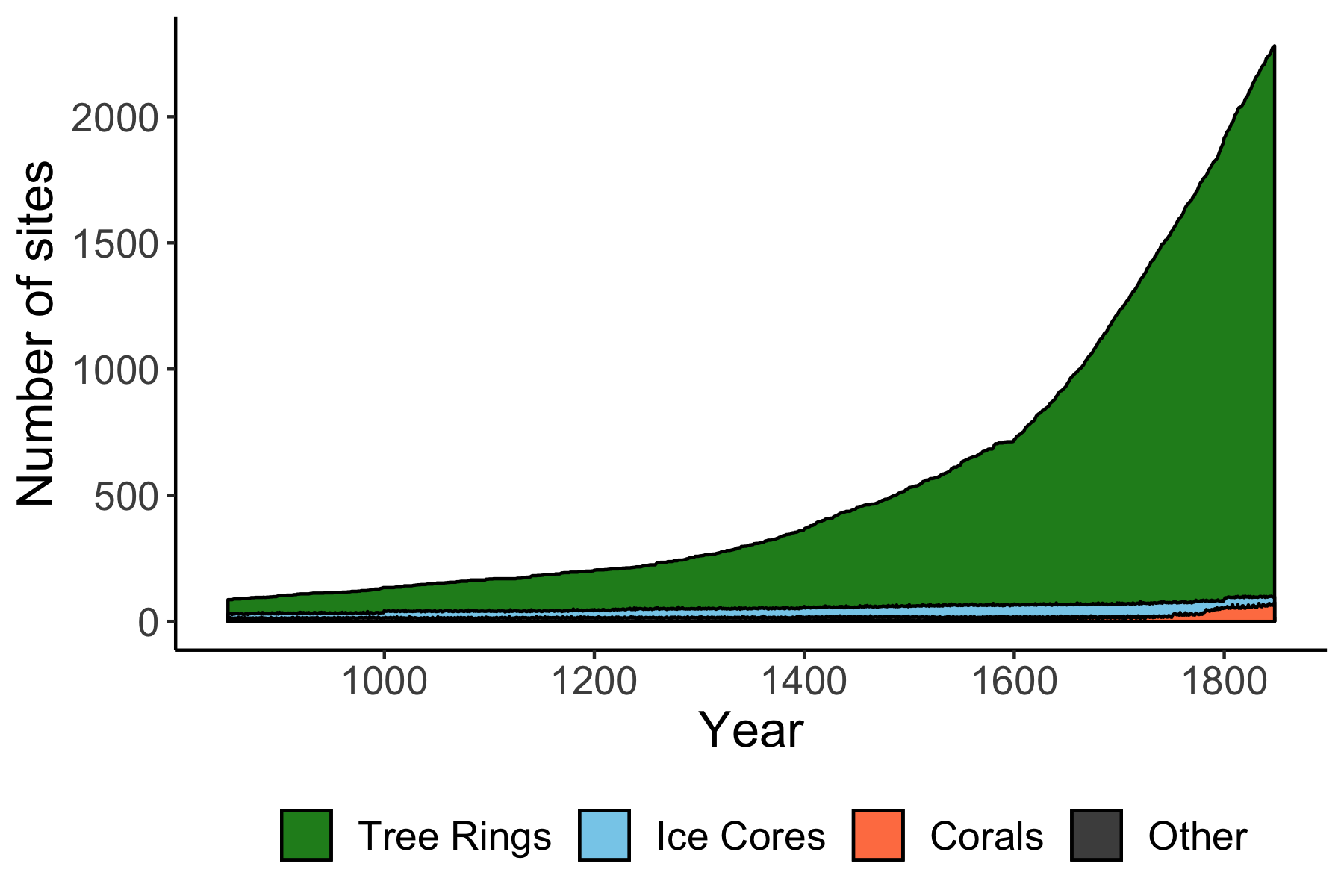}
    \caption{Temporal distribution of proxies by the three largest categories (tree rings, ice cores, and corals). There are total of 2468 proxies used over this interval, with the vast majority being tree rings.}
    \label{fig:temp_dist}
\end{figure}

A collection of modeled climate fields from the CESM simulation are used to form the background state in the DA scheme. For the purpose of our analysis herein, we will specifically use the modeled and reconstructed 2-meter surface temperature fields. The temperature fields are processed from the native model output by annual averages and spatially discretizing onto a $2^{\circ}$ latitude and longitude grid ($144 \times 96$ grid points). Annual in this context is defined as the interval between April and March of the following year, thus yielding 998 such climatological years to be used for the background ensemble. Because of the large data files produced by PHYDA, we only used a 100 member sub-ensemble, randomly drawn from the original 998 member ensemble, for our analyses. The final processed background state, therefore, consists of 100 spatial fields, each observed on the same $144 \times 96$ grid points.

The 998 analysis states are derived from the background state by using DA to incorporate temporally available proxy information during each year of reconstruction (See Figure \ref{fig:temp_dist}). Each analysis state is also a 100 member ensemble of 2m surface temperature fields discretized to the same $2^{\circ}$ latitude and longitude grid as the background ensemble. 
Quantifying the influence that proxies have in the analysis states is quite challenging due to their small individual effect sizes, and the fact that they can affect higher order structures of the data beyond the mean and variance. Identifying the full effect of the proxies would, therefore, require testing for distributional changes.



\begin{figure}
    \includegraphics[width=\textwidth]{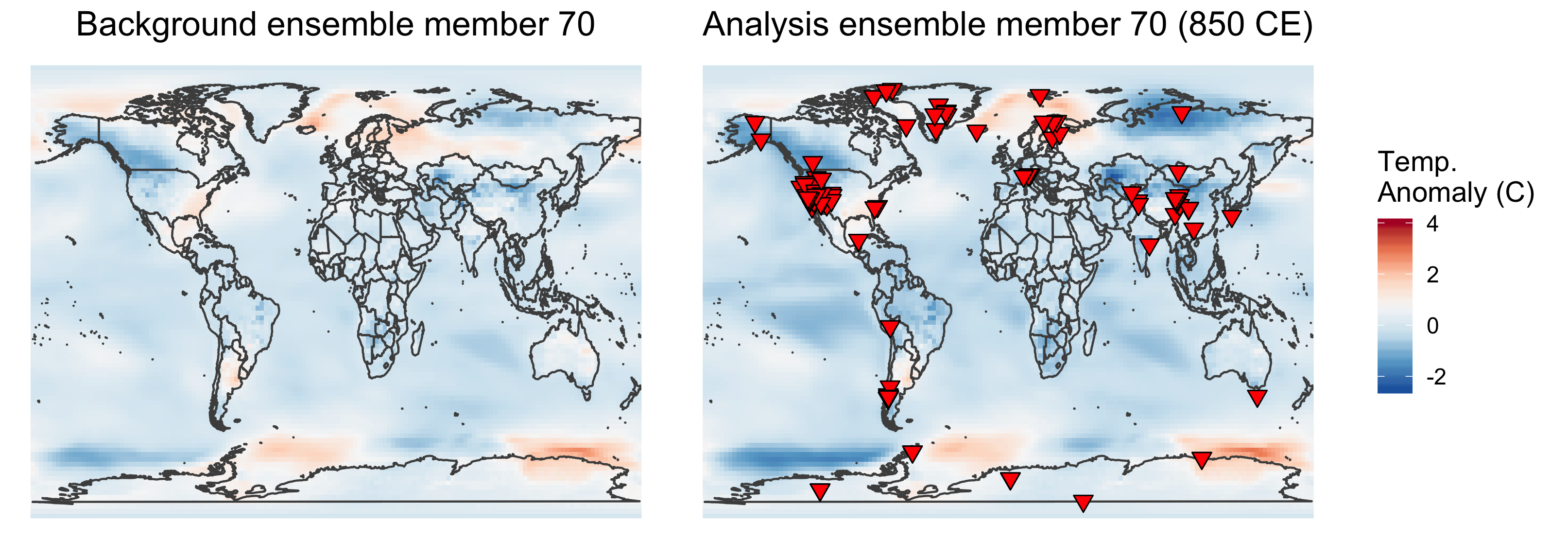}
    \caption{Temperature anomalies for a single background and analysis ensemble member with respect to the background mean field. Left panel is from the CESM simulation run while the right panel is from PHYDA during 850 CE. Red triangles indicate the locations of proxies available in 850 CE.}
  \label{fig:example}
\end{figure}

\section{Statistical Solution} \label{sec:solution}
We first formulate our scientific problem into a hypothesis testing, then introduce the integrated Tukey depth and propose our test statistic, followed by a discussion of the asymptotic distribution of our test statistics under the null hypothesis. 

\subsection{Formulation of evaluating proxy influence} \label{sec:eval} 
Let $X$ and $Y_t$ respectively represent the ensemble in the background state and the ensemble in the analysis state at time $t$ in PHYDA.  Under the assimilation design, the proxies at time $t$ are the only contributors to the differences between the two sets of ensembles. Our goal is to define and quantify the differences between $X$ and $Y_t$ each year in order to assess the proxy influence.

The amount that proxies impact the analysis states depends on many factors including the proxy type (e.g., tree ring, ice core, coral), where proxies were collected, and the interval over which the proxies were observed \citep{steiger2018PHYDA}. As shown in Fig. \ref{fig:example}, the effects from proxies may be small and thinly diffused over a non-contiguous area due to spatial correlations and teleconnections. In fact, most of the induced mean differences generally fall within the natural variation of the background fields. The most comprehensive approach to test for the proxy's cumulative influence is, therefore, to test for changes in the distributions of $X$ and $Y_t$. We thus formulate our problem into the following hypotheses: 
\begin{align}\label{eq:test1}
    H_0: X \stackrel{D}{=} Y_t \ \ \text{ v.s. }  \ H_A: X \stackrel{D}{\neq} Y_t,
\end{align}
where $\stackrel{D}{=}$ means equality in distribution.
In addition to the outcome of these hypothesis tests at each time $t$, we are also equally interested in the pattern of those outcomes as $t$ increases. Over time, the amount of proxy information available for reconstruction increases, while the background ensemble stays the same. We therefore might expect that the divergence between the background and the analysis distributions will increase over time accordingly if the proxies are having their due influence.

Under the functional data analysis regime, we assume that the observed data are generated from continuous functions combined with additive noise, instead of from a spatially correlated stochastic processes. In this framework, each ensemble member represents a single observation over a spatial domain where $144 \times 96$ grid points are embedded. This distinction allows us to consider each ensemble member as an \textit{i.i.d} realization of a stochastic process in a functional space. 


We develop our test statistic for the testing problem (\ref{eq:test1}) at any given $t$ in a general context.   
For ease of notation, we suppress $t$ from $Y_t$. 
Let $X = \{X_i(s)\}_{i = 1}^n$ and $Y = \{Y_j(s)\}_{j = 1}^m$, where $s \in D$ and $D$ is a compact subspace of $\mathbbm{R}^p$. Without loss of generality, let $D$ be $[0, 1]^p$ and let each functional datum be observed at the same locations in $[0, 1]^p$.  We assume that each function $X_i$ and $Y_j$ is a univariate continuous function on the domain $[0, 1]^p$, i.e. $X_i : [0, 1]^p \mapsto \mathbbm{R} \hbox{ for } i \in 1,\ldots,n$; $Y_j : [0, 1]^p \mapsto \mathbbm{R} \hbox{ for } j \in 1,\ldots,m.$
In other words, each $X_i$ (or $Y_j$) is an element of the class of univariate continuous functions on $[0, 1]^p$, denoted by $C[0, 1]^p$.
Specific to our data, we have $p = 2$ and $X_i(s)$ and $Y_j(s)$ respectively represent the $i$th background state and the $j$th analysis state at location $s$.

Let $P$ and $Q$ be two absolutely continuous distributions on $C[0,1]^2$ and suppose each $X_i \sim P$ and each $Y_j \sim Q$. We are interested in testing if the functional data in $X$ and in $Y$ follow the same distribution, so (\ref{eq:test1}) is equivalent to the hypotheses,
\begin{align}\label{eq:test2}
    H_0: P = Q; \ \ \text{ v.s. } \ H_A: P \neq Q,
\end{align}
for any given $t$. 
We will use functional data depth to construct a two sample Kolmogorov-Smirnov-type test. Other distribution free tests such as the Anderson-Darling or Cramer-Von Mises test could equally have been applied. We chose Kolmogorov-Smirnov for its convenient asymptotic form and its ubiquity in testing distributions.

\subsection{Integrated Tukey depth} \label{depth}
Data depth is a statistical concept for quantifying the centrality or ``depth'' of the observed data points with respect to a reference distribution. The closer an observation is to the center of the distribution, the higher its depth value should be to indicate its centrality. As the reference distribution is typically unknown, the depth of an observation has to be estimated via an empirical notion of data depth. Many notions of data depth for functional data have been developed including
the integrated band depth \citep{banddepth}, extremal depth \citep{naveen}, and various integrated univariate depths \citep{muniz}. Each of these depth functions has its own strengths and weaknesses but none dominates the others in all aspects, see \citet{cuevas} and \citet{nagy} for a review. We chose the integrated Tukey depth as the basis of our test for its simplicity, robustness, computational tractability, and highly desirable theoretical properties. 

Integrated depths are a well studied class of functional data depth measures that were first introduced by \citet{muniz} and then studied extensively by \citet{cuevas} and \citet{nagy}.
To define an integrated depth function, a univariate depth function is first defined over a collection of one dimensional ``projections" of the data which often refers to the observed values of the functions at each location $s \in D$. The univariate depth is then integrated over these projections to yield the integrated depth. Among all the univariate depths, the Tukey depth and the simplicial depth are perhaps the two most popular ones. We opted to use the Tukey depth but the simplicial depth would have been equally effective because the orderings they induce are nearly identical.

The integrated Tukey depth is defined as follows. Let $P$ be a distribution for $X \in C[0, 1]^p$, and let $P_s$ be the marginal distribution of $P$ at $s \in [0, 1]^p$. The univariate Tukey depth of $X(s) = x(s)$ with respect to $P_s$ is
\[
D(x(s), P_s) = 1 - |1 - 2P_s(x(s))|,
\]
and the integrated Tukey depth of $X=x$ with respect to $P$ is
\[
D(x, P) = \int_{[0, 1]^p}D(x(s), P_s)ds.
\]

To ensure that this depth function is proper, we refer to the criteria proposed by \citet{zuo} and \citet{mosler2012general}. In \citet{nagy} it was shown that the integrated Tukey depth satisfies translation invariance, function scale invariance, measure-preserving rearrangement invariance, maximality at the center, continuity, and quasi-concavity of the induced level sets.
They also demonstrated strong universal consistency and weak uniform consistency, which assure that the integrated Tukey depth behaves well and asymptotically converges to its population counterpart under regularity conditions.

\subsection{Test statistic} \label{test}
We propose a test statistic $KD(X, Y)$, called the Kolmogorov Depth ($KD$) statistic, for our hypothesis testing problem (\ref{eq:test2}) based on the integrated Tukey depth. The $KD$ statistic measures the outlyingness of a sample $X \sim P$ from the distribution $Q$ as well as the outlyingness of a sample $Y \sim Q$ from the distribution $P$. It takes the maximum of the two outlyingness measures as its value. This way we can correctly detect differences between $P$ and $Q$ even when they do not appear mutually outlying from each other under data depth.
For example, if one of the distributions is nested inside the other then the nested distribution will not appear outlying to the other distribution.

Denote $P_n$ as the empirical estimate of $P$ based on the sample $X = \{X_1=x_1,\dots,X_n=x_n\}$ and $Q_m$ the empirical estimate of $Q$ based on $Y = \{Y_1=y_1,\dots,Y_m=y_m\}$. We start by considering $P_n$ fixed and aim to measure the outlyingness of $Q_m$ over $P_n$. To do this we first define the following two empirical measures for any given $x_k \in X$:
\begin{align}
  \widehat{F}_X(x_k) &= \frac{1}{n}\sum_{i=1}^n \mathbbm{1}(D(x_i, P_n) \leq D(x_k, P_n)) \label{eq:Fn}\\
    \widehat{G}_Y(x_k) &= \frac{1}{m}\sum_{j=1}^m \mathbbm{1}(D(y_j, P_n) \leq D(x_k, P_n)) \label{eq:Gn}.
\end{align}
Essentially, $\widehat{F}_X(x_k)$ is a rescaling of $D(x_k, P_n)$ to its standardized rank, i.e. $\widehat{F}_X(x_k) = 1/n$ if $D(x_k, P_n)$ is the smallest, $\widehat{F}_X(x_k) = 2/n$ if $D(x_k, P_n)$ is the second smallest, and so on.
It acts as the empirical cumulative distribution function of $D(x_k, P_n) \hbox{ for } k \in 1,\dots,n$ evaluated at itself and thus follows a discrete uniform distribution.
The second quantity $\widehat{G}_Y(x_k)$ can be considered as the empirical cumulative distribution function of $D(y_k, P_n)$ evaluated at $D(x_k, P_n)$. Under $H_0$ in (\ref{eq:test2}), $\widehat{G}_Y$ should be approximately uniform, so a deviation of $\widehat{G}_Y$ from the uniform distribution indicates an outlyingness of $Q_m$ from $P_n$. The introduction of $ \widehat{F}_X(x_k)$ and $\widehat{G}_Y(x_k)$ allows us to reduce the problem of comparing two sets of random fields to assessing the difference in distribution between two sets of random variables, $\widehat{F}_X(x_k)$ and $\widehat{G}_Y(x_k)$ for $k=1,\ldots, n$. The latter can be naturally quantified using the Kolmogorov distance over the set $X$:
\begin{equation}\label{eq:KPn}
K_{P_n}(X, Y) = \max_{x_k \in X}|\widehat{F}_X(x_k) - \widehat{G}_Y(x_k)|.
\end{equation}

To measure the outlyingness of $P_n$ over $Q_m$ we now fix $Q_m$ rather than $P_n$. Following the same scheme, we define the two empirical measures for any given $y_k \in Y$ as:
\begin{align*}
    \widetilde{F}_X(y_k) &= \frac{1}{n}\sum_{i=1}^n \mathbbm{1}(D(x_i, Q_m) \leq D(y_k, Q_m)) \\
    \widetilde{G}_Y(y_k) &= \frac{1}{m}\sum_{j=1}^m \mathbbm{1}(D(y_j, Q_m) \leq D(y_k, Q_m)).
\end{align*}
These two quantities exactly mirror $\widehat{F}_X$ and $\widehat{G}_Y$ except that now $\widetilde{G}_Y$ is uniform on the depth values $D(y_k, Q_m)$, for $k \in 1,\dots,m$, and $\widetilde{F}_X$ is the indicator for the outlyingness of $P_n$ from $Q_m$. We again take the Kolmogorov distance, but now over the set $Y$, as the measure of outlyingness
\[
K_{Q_m} (X, Y) = \max_{y_k \in Y}|\widetilde{F}_X(y_k) - \widetilde{G}_Y(y_k)|.
\]

We define the overall test statistic $KD$ by taking the maximum of the two distances:
\begin{equation}\label{eq:ts}
KD(X, Y) = \max\{K_{P_n}(X, Y) \text{, } K_{Q_m}(X, Y)\}.
\end{equation}
The test statistic $KD$ attains a level of symmetry by making the test invariant to the reference distribution. It is strictly non-negative and it equals 0 only under $H_0$ in the hypothesis (\ref{eq:test2}). Thus the originally stated hypothesis (\ref{eq:test1}) can be tested by evaluating whether $KD$ is significantly greater than 0. 

One major difference between our test statistic $KD$ and the $QI$ in \cite{quality} is that our test does not depend on a reference distribution while $QI$ requires one of the samples to be used as the reference. Our test computes the outlyingness of two samples from each other and aggregates the results into one single test. This is a more efficient use of the two samples and enables $KD$ to detect a larger range of alternative hypotheses, such as the nesting situation mentioned above. We discuss the critical values of $KD$ in the following section.

\subsection{Computing critical values} \label{critical}
Deriving the asymptotic distribution of $KD$ is nontrivial because $KD$ explicitly depends on two non \textit{i.i.d.} processes, $D(x_i, P_n)$ and $D(y_j, Q_m)$. This renders standard results on the Kolmogorov-Smirnov test inapplicable. Nevertheless, we conjecture without formal proof that $KD$ either follows the same limiting distribution as the regular Kolmogorov-Smirnov two sample statistic,  i.e.
\[
\sqrt{\frac{nm}{n+m}} KD \xrightarrow{D} K',
\]
where
\[
P(K' < t) = 1 - 2\sum_{j=1}^{\infty}(-1)^je^{-2j^2t^2},
\]
or converges to a distribution that can be closely approximated by $K'$. Although we are unable to prove this result in its full generality, we consider two special cases below and show that both conform to the conjecture of converging to $K'$. Our extensive simulation studies in Section \ref{sims} demonstrate convergence in the general case.


We first consider a special case where $P$ is known and we are interested in testing if $Y_j\sim P$ for $j=1,\ldots, m$. In this case, $\widehat{F}_X(x_k)$ in (\ref{eq:Fn}) becomes the uniform$[0, 1]$ distribution at $D(x_k, P)\in [0,1]$. Then $K_{P_n}(X, Y)$ in (\ref{eq:KPn}), which is the test statistic in this special case, reduces to 
\[
K_{P}(Y) = \sup_{x_k}|D(x_k, P) - \widehat{G}_Y(x_k)|. 
\]
Because $\widehat{G}_Y(x_k)$ is an empirical distribution of the $i.i.d.$ random variables $\{D(y_1, P), \dots, D(y_m, P) \}$ at $D(x_k, P)$, $K_{P}(Y)$ is exactly the one sample Kolmogorov-Smirnov statistic for testing the uniformity of $\widehat{G}_Y(x_k)$. Therefore,
\[
\sqrt{m} K_P(Y) \xrightarrow{D} K'.
\]


We further consider another special case where $P$ and $Q$ are both unknown but with either $n \gg m$ or $m \gg n$. We can show that $K_{P_n}(X, Y)$ (or $K_{Q_m} (X, Y)$) converges to the Kolmogorov distribution under $n \gg m$ ($m \gg n$).  
We encapsulate this result in the following proposition.
\begin{proposition} \label{asymptotic}
Suppose that $n \gg m$, then under the null hypothesis,
\[
\sqrt{\frac{nm}{n+m}} K_{P_n}(X, Y) \xrightarrow{D} K',
\]
where $K'$ follows the Kolmogorov distribution.
\end{proposition}
The proof is deferred to the Appendix.

Generalizing the results of these special cases is challenging. This issue was also noted in \citet{quality} where the authors conjectured that their two sample $QI$ asymptotically followed a normal distribution, as its one sample version does. Their conjecture was only later proven in \citet{xuming} after substantial theoretical development. 
The techniques that emerged from the proof in \citet{xuming} relied heavily on $QI$ being an expectation, making them largely inapplicable to our context involving suprema. Proving the conjecture would require the development of advanced theoretical machinery that can accommodate the complex dependence nature of the distribution functions of the depth measures. We leave this problem open for independent theoretical research in the future.

In lieu of the proposed asymptotic distribution, we may consider using permutations to find critical values for $KD$ \citep{good2013permutation}. Permutation works well for small samples or sparsely observed functions, but it quickly becomes computationally infeasible on large volumes of data, such as our reconstruction data. For this reason, the conjectured Kolmogorov distribution is more appealing in practice.

\section{Simulation Study} \label{sims}
Simulation studies are conducted to assess the convergence of $KD$ to $K'$, and the size and power of the test. 
Each of these properties is evaluated using two-dimensional functional data because our main application considers ensembles of spatial fields. All functional data in the simulation are generated from Gaussian random processes with the Mat\'ern covariance function \citep{stein2012interpolation}, 
\[
C(x, x') = \frac{\sigma \sqrt{\pi} r^{2\nu}}{2^{\nu-1}\Gamma(\nu + 1/2)} \left(\frac{\|x - x'\|}{r} \right)^\nu K_{\nu} \left(\frac{\|x - x'\|}{r} \right),
\]
where $\Gamma$ is the Gamma function, $K_{\nu}$ is a modified Bessel function, $\sigma$ is the marginal variance of the random process, and $r$ and $\nu$ are two nonegative parameters called range and smoothness. The range parameter, $r$, governs how quickly the correlation decays between points. 
The smoothness parameter, $\nu$, determines how smooth sampled functions are in terms of their differentiability.

In each simulation, we consider the sample $X$ as the baseline and $Y$ as the sample to be varied. For the size and convergence simulations the marginal variance $\sigma$ will always be set to 1, while $r$ and $\nu$ will be allowed to vary. For the power simulations $\mu$, $\sigma$, $r$, and $\nu$ will all be allowed to vary.

\subsection{Convergence} \label{sec:conv}
We use simulations to validate the conjectured asymptotic Kolmogorov distribution of our test statistic (\ref{eq:ts}) under the null hypothesis. The main idea is to evaluate how well the permutation distribution of the test statistic is approximated by the Kolmogorov distribution, even at moderate sample sizes. Functional data $X = \{X_1,\dots,X_n\}$ and $Y = \{Y_1,\dots,Y_n\}$ are each generated with mean, $\mu = 0$, and standard deviation, $\sigma = 1$, on the spatial domain $[0, 1] \times [0, 1]$. Because the integrated Tukey depth is invariant to the location and scale of functional data, we only vary the range and smoothness of the covariance function: 0.2, 0.3, 0.4, 0.5 and 0.5, 1.0, 1.5, respectively. The number of replicates, $n$, in each sample is also varied between 25, 50, 75, 100. We also considered the unbalanced sample size case by fixing the number of replicates in $Y$ to be 75 and allowing the number in $X$ to vary between 25, 50, 75, 100. The results were nearly identical as to those when the sample sizes were balanced so only the balanced case is presented here. 

\begin{figure}
    \begin{center}
    \includegraphics[width=0.80\textwidth,valign=c]{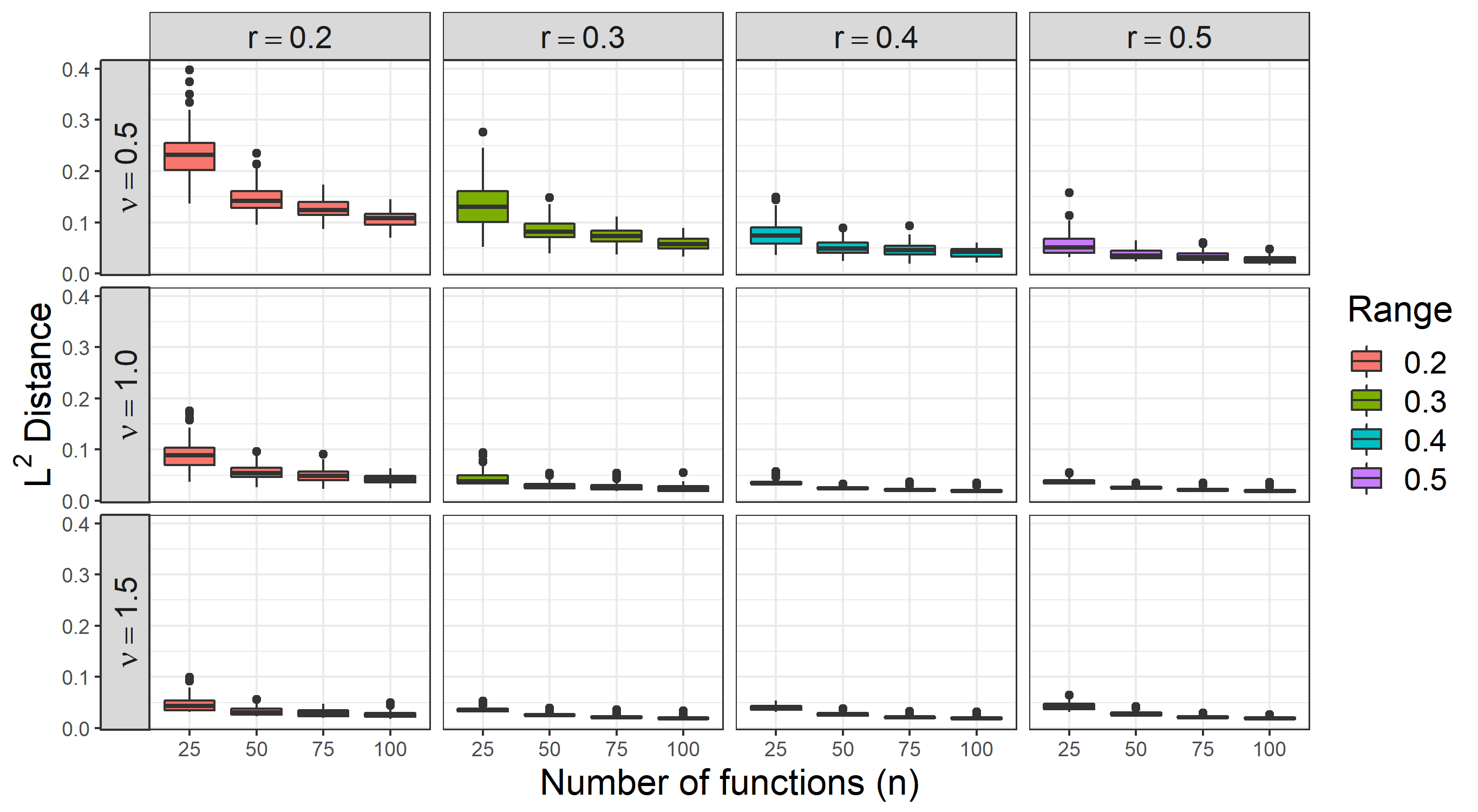}
    \caption{$\Ltwo$ distance between the permutation distribution and the Kolmogorov distribution under 12 different range, $r$, and smoothness, $\nu$, settings.}
    \label{fig:l2}
    \end{center}
\end{figure}

\begin{figure}
    \begin{center}
    \includegraphics[width=0.80\textwidth,valign=c]{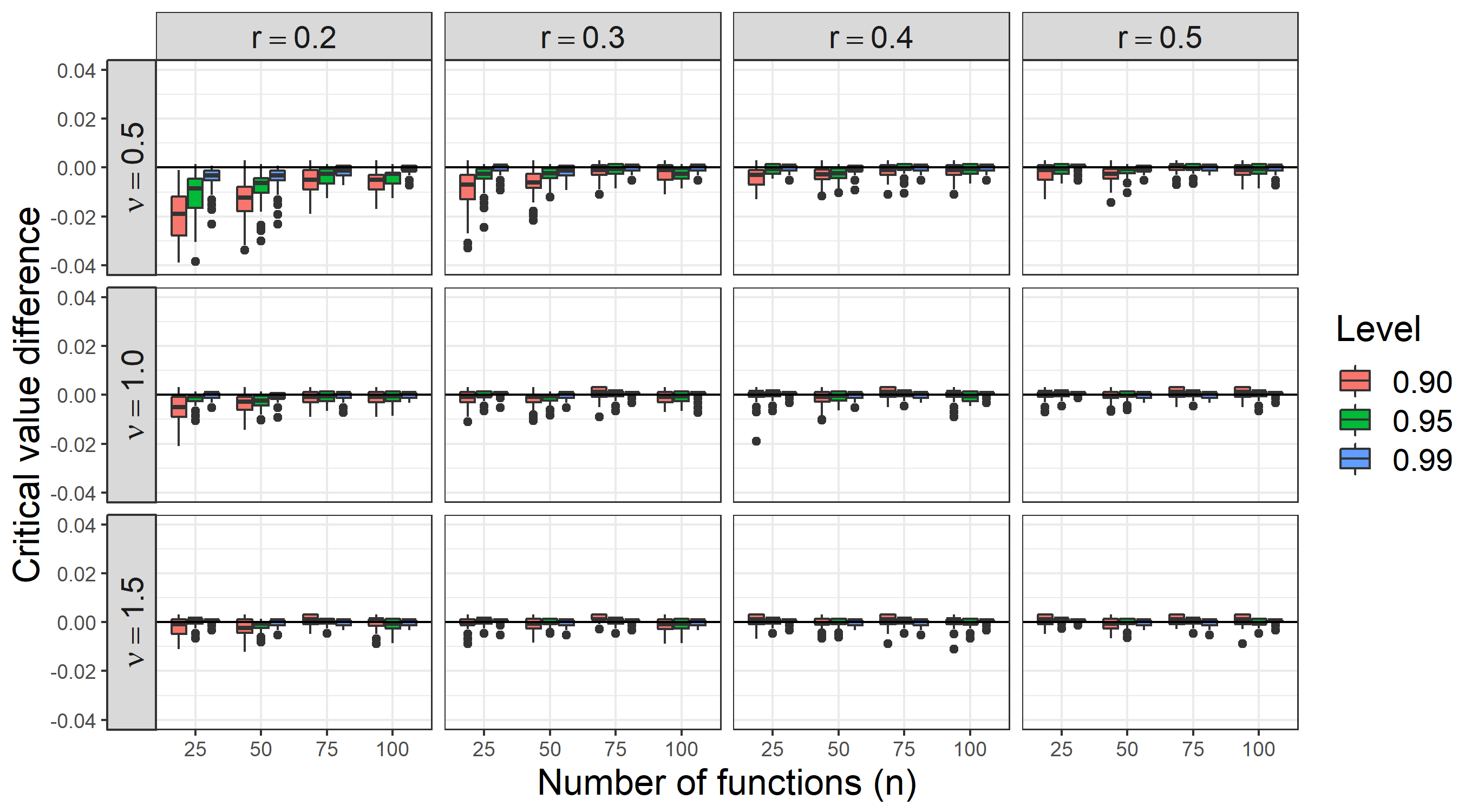}
    \caption{Kolmogorov critical values minus permutation critical values at three common test levels: 0.90, 0.95. 0.99 under 12 different range, $r$, and smoothness, $\nu$, settings.}
    \label{fig:crit}
    \end{center}
\end{figure}

The permutation distribution was constructed by recomputing $KD$ on 500 permutations of the generated $X$ and $Y$ samples. We then calculated the $\Ltwo$ distance between the permutation distribution and the Kolmogorov distribution and the difference between critical values derived from the permutation and Kolmogorov distribution at three common significance levels: 0.01, 0.05, and 0.10. Due to the computational cost of constructing permutation distributions we ran 100 simulations for each combination of $r$, $\nu$, and $n$ to obtain the boxplots in Figures \ref{fig:l2} and \ref{fig:crit}.

Figure \ref{fig:l2} demonstrates convergence of the permutation distribution to Kolmogorov in $\Ltwo$. For even small sample sizes, such as $n = 25$, the distance between the two distributions is already vanishingly small for smooth data ($r \geq 0.3$ and $\nu \geq 1.0$).
The largest deviations are only observed when both the range and smoothness are small, $r < 0.3$ and $\nu < 1.0$. This is typically not an issue in practice because functional data are generally preprocessed with a smoothing step; effectively increasing $\nu$ and $r$. In all cases the $\Ltwo$ norm decreases rapidly with an increasing sample size such that the convergence even applies to unprocessed noisy data if the sample sizes are large enough.

Figure \ref{fig:crit} evaluates the convergence of the two sets of critical values at the significance levels 0.01, 0.05, and 0.10.
This figure shows that testing decisions reached under the asymptotic Kolmogorov distribution are generally not biased away from decisions reached under the permutation decision. Again, a sufficient amount of smoothness ($r \geq 0.3$ or $\nu \geq 1.0$) is required to have well-behaved critical values. If the data are not sufficiently smooth then the Kolmogorov distribution tends to have smaller critical values than the corresponding permutation distribution. The size will therefore be slightly inflated by using Kolmogorov and so the permutation distribution should be preferred when computationally feasible. Once a sufficient level of smoothness has been reached, in this case $r \geq 0.3$ or $\nu \geq 1.0$, the critical values of the permutation distribution become highly comparable with the Kolmogorov distribution. The observed differences are minuscule such that any decision reached using the Kolmogorov distribution is likely to be the same as if the permutation distribution were used. With noisy raw data a sufficient number of samples ($n \geq 100$) can begin to compensate for a lack of smoothness or correlation.

\subsection{Size and power} \label{size}
Using the same data generating process as in Section \ref{sec:conv}, we evaluate the size of our test using critical values from the asymptotic Kolmogorov distribution and compare our size to the $QI$ test. 
Again only $r$, $\nu$, and the two sample sizes $n$ and $m$ will be varied. The size under each combination of $r$, $\nu$, $n$, and $m$ was estimated using 2000 simulations; the results of which are presented in Table \ref{tab:size}.

\begin{table}
\scriptsize
\centering
\begin{tabular}{cc|cccc|cccc|cccc}
  \hline
  \multicolumn{2}{c}{} & \multicolumn{4}{|c}{$\nu = 0.5$}& \multicolumn{4}{|c}{$\nu = 1.0$} & \multicolumn{4}{|c}{$\nu = 1.5$}\\ 
 n & m & r = 0.2 & 0.3 & 0.4 & 0.5 & 0.2 & 0.3 & 0.4 & 0.5 & 0.2 & 0.3 & 0.4 & 0.5 \\ 
 \hline
50 & 50 & 0.15 & 0.09 & 0.08 & 0.06 & 0.07 & 0.06 & 0.04 & 0.05 & 0.06 & 0.05 & 0.05 & 0.06 \\ 
 & & (0.24) & (0.17) & (0.15) & (0.13) & (0.13) & (0.13) & (0.11) & (0.09) & (0.13) & (0.10) & (0.09) & (0.08) \\ 
50 & 100 & 0.13 & 0.10 & 0.07 & 0.06 & 0.07 & 0.06 & 0.04 & 0.05 & 0.06 & 0.06 & 0.05 & 0.06 \\ 
 & & (0.28) & (0.21) & (0.17) & (0.13) & (0.17) & (0.14) & (0.11) & (0.10) & (0.14) & (0.11) & (0.10) & (0.09) \\ 
50 & 200 & 0.16 & 0.10 & 0.07 & 0.07 & 0.08 & 0.06 & 0.06 & 0.05 & 0.07 & 0.05 & 0.06 & 0.05 \\ 
 & & (0.32) & (0.26) & (0.19) & (0.16) & (0.20) & (0.13) & (0.13) & (0.10) & (0.16) & (0.11) & (0.11) & (0.09) \\ 
50 & 300 & 0.14 & 0.09 & 0.08 & 0.06 & 0.07 & 0.06 & 0.06 & 0.05 & 0.06 & 0.06 & 0.05 & 0.04 \\ 
 & & (0.39) & (0.23) & (0.19) & (0.16) & (0.19) & (0.13) & (0.13) & (0.11) & (0.16) & (0.12) & (0.11) & (0.09) \\ \hline
100 & 50 & 0.15 & 0.10 & 0.07 & 0.06 & 0.09 & 0.07 & 0.06 & 0.05 & 0.06 & 0.05 & 0.05 & 0.04 \\ 
 & & (0.15) & (0.11) & (0.09) & (0.08) & (0.10) & (0.08) & (0.07) & (0.07) & (0.09) & (0.07) & (0.07) & (0.06) \\ 
100 & 100 & 0.10 & 0.07 & 0.07 & 0.05 & 0.06 & 0.05 & 0.04 & 0.05 & 0.06 & 0.04 & 0.04 & 0.04 \\ 
 & & (0.18) & (0.13) & (0.12) & (0.11) & (0.11) & (0.09) & (0.08) & (0.08) & (0.11) & (0.08) & (0.08) & (0.07) \\ 
100 & 200 & 0.11 & 0.07 & 0.07 & 0.06 & 0.06 & 0.05 & 0.06 & 0.05 & 0.06 & 0.05 & 0.05 & 0.05 \\ 
 & & (0.22) & (0.15) & (0.11) & (0.12) & (0.12) & (0.10) & (0.10) & (0.10) & (0.10) & (0.08) & (0.08) & (0.08) \\ 
100 & 300 & 0.11 & 0.06 & 0.06 & 0.06 & 0.07 & 0.06 & 0.05 & 0.05 & 0.06 & 0.05 & 0.05 & 0.04 \\ 
 & & (0.22) & (0.16) & (0.14) & (0.10) & (0.15) & (0.11) & (0.08) & (0.08) & (0.11) & (0.09) & (0.09) & (0.07) \\ \hline
200 & 50 & 0.15 & 0.10 & 0.07 & 0.08 & 0.08 & 0.06 & 0.06 & 0.06 & 0.06 & 0.06 & 0.05 & 0.05 \\ 
 & & (0.09) & (0.08) & (0.07) & (0.07) & (0.07) & (0.07) & (0.06) & (0.06) & (0.07) & (0.07) & (0.06) & (0.06) \\ 
200 & 100 & 0.10 & 0.08 & 0.06 & 0.06 & 0.07 & 0.06 & 0.05 & 0.05 & 0.06 & 0.04 & 0.05 & 0.04 \\ 
 & & (0.11) & (0.09) & (0.07) & (0.06) & (0.09) & (0.07) & (0.07) & (0.06) & (0.08) & (0.06) & (0.07) & (0.06) \\ 
200 & 200 & 0.08 & 0.06 & 0.05 & 0.05 & 0.06 & 0.05 & 0.04 & 0.05 & 0.05 & 0.05 & 0.06 & 0.04 \\ 
 & & (0.13) & (0.10) & (0.08) & (0.08) & (0.09) & (0.08) & (0.08) & (0.07) & (0.08) & (0.08) & (0.07) & (0.07) \\ 
200 & 300 & 0.08 & 0.07 & 0.06 & 0.06 & 0.06 & 0.06 & 0.05 & 0.05 & 0.06 & 0.05 & 0.06 & 0.06 \\ 
 & & (0.12) & (0.11) & (0.10) & (0.09) & (0.09) & (0.07) & (0.07) & (0.07) & (0.07) & (0.07) & (0.07) & (0.07) \\ \hline
300 & 50 & 0.14 & 0.08 & 0.06 & 0.06 & 0.07 & 0.06 & 0.06 & 0.05 & 0.06 & 0.06 & 0.05 & 0.05 \\ 
 & & (0.07) & (0.06) & (0.06) & (0.05) & (0.06) & (0.07) & (0.06) & (0.06) & (0.06) & (0.07) & (0.06) & (0.06) \\ 
300 & 100 & 0.10 & 0.06 & 0.05 & 0.05 & 0.06 & 0.06 & 0.04 & 0.05 & 0.05 & 0.05 & 0.04 & 0.05 \\ 
 & & (0.09) & (0.08) & (0.07) & (0.06) & (0.06) & (0.07) & (0.06) & (0.06) & (0.06) & (0.06) & (0.07) & (0.05) \\ 
300 & 200 & 0.09 & 0.06 & 0.06 & 0.07 & 0.07 & 0.05 & 0.04 & 0.04 & 0.06 & 0.05 & 0.05 & 0.05 \\ 
 & & (0.10) & (0.07) & (0.07) & (0.08) & (0.07) & (0.07) & (0.06) & (0.06) & (0.08) & (0.06) & (0.07) & (0.07) \\ 
300 & 300 & 0.07 & 0.06 & 0.06 & 0.06 & 0.05 & 0.05 & 0.05 & 0.04 & 0.05 & 0.05 & 0.06 & 0.05 \\ 
 & & (0.11) & (0.09) & (0.09) & (0.08) & (0.08) & (0.08) & (0.07) & (0.06) & (0.06) & (0.07) & (0.06) & (0.07) \\ 
   \hline
\end{tabular}

\caption{Sizes of $KD$ and $QI$ (in parenthesis) under 12 combinations of range, $r$, and smoothness, $\nu$, and 16 combinations of sample sizes, $n$ and $m$, for $X$ and $Y$ respectively.}
\label{tab:size}
\end{table}

Our simulations show that for even small samples, such as $n = m = 50$, our test can control the size near the prescribed level if the range or smoothness is sufficiently high; that is $r \geq 0.4$ or $\nu \geq 1$.
Smoothness and range are in fact more important for controlling size than the number of replicates. Under the noisiest setting, $r = 0.2$ and $\nu = 0.5$, the lowest attained size (0.07) occurs when $n = m = 300$. This is an only moderate improvement over the size (0.15) when $n = m = 50$, and is still above the nominal level. If instead the number of replicates were fixed at $n = m = 50$ but the range and smoothness increased to either $0.5$ and $1.0$, respectively, then the size is controlled at the nominal level and thereafter.
As with the convergence simulations though, these minimal smoothness conditions are not all that impactful in practice because functions are typically smoothed before analysis.
Moreover, the sizes are stable at the nominal level once the range exceeds a threshold between 0.3 and 0.4 or $\nu \geq 1.0$ for the spatial domain $[0, 1] \times [0, 1]$.  The $QI$ test appears to inflate the size in nearly all cases compared to our test.

We compare the power of our test and the $QI$ test in detecting changes in the four parameters $\mu$, $\sigma$, $r$, and $\nu$ that govern the underlying Gaussian process in our data generation. We set the number of replicates to $n = 100$ and $m = 50$ for the two samples $X$ and $Y$, respectively, and sample the functional observations in $X$ from  a Gaussian process with $r = 0.4$, $\nu = 1$, $\mu = 0$, and $\sigma = 1$. This setup ensures that the sizes of $KD$ and $QI$ are similar (see Table \ref{tab:size}) so that their power functions are comparable. To generate samples of $Y$ we let each of the parameters in $Y$ vary around the parameter values in $X$. The mean, $\mu$, was set from -1 to 1 in 0.1 increments, $\sigma$ was set between 0.1 and 2 in 0.05 increments, $r$ from 0.05 to 1 in 0.05 increments, and $\nu$ from 0.1 to 2 in 0.1 increments. This gave a total of 96 alternative models because the parameters were varied individually. The power of $KD$ and $QI$ were then calculated under each of these alternative models using 2000 simulations each.

\begin{figure}
    \begin{center}
    \includegraphics[width=\textwidth,valign=c]{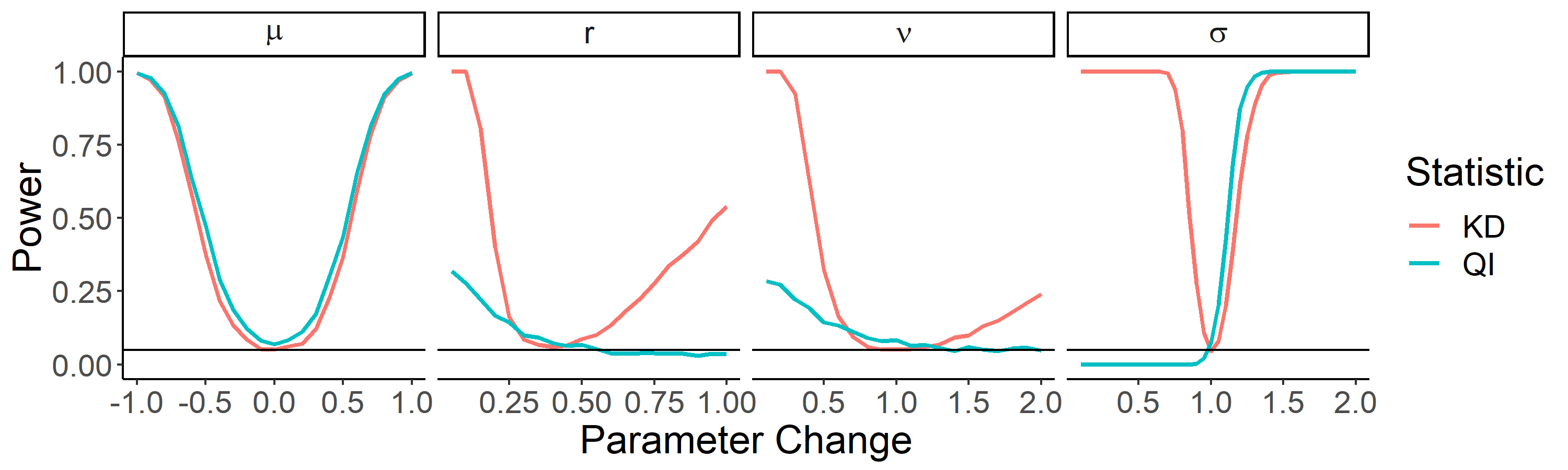}
    \caption{Power of $KD$ and $QI$ in detecting changes in the four parameters in the Gaussian process. Mean, Range and smoothness are presented as shifts of parameters in $Y$ from $X$. Standard deviation is presented as a multiple of standard deviation in $X$.}
    \label{fig:power_const}
    \end{center}
\end{figure}

Figure \ref{fig:power_const} shows the empirical power functions for both $KD$ and $QI$ on each parameter. Both tests are almost equally powerful in detecting mean changes and increases in standard deviation. However, $KD$ shows a strong improvement over $QI$ in detecting changes in range, smoothness, and decreases in standard deviation. Two caveats about the power functions should be noted. The first is that there is a slight advantage to $QI$ in the testing of mean, range and smoothness because $QI$ still appears to slightly inflate size, which can be seen by observing its power function at the null value of these three parameters. The second caveat is that $QI$ was not designed to detect decreases in standard deviation because the application for which it was designed found a drop in standard deviation desirable.

Further simulation results comparing against the Functional Anderon-Darling ($FAD$) and the Rank based Band Depth Test ($BAND$) are available in the supplement. While $FAD$ shows considerable power in detecting mean changes it falls short of the other methods for detecting variance changes. On average our method maintained the highest power across the different parameters on average.


The situation where the mean or variance is shifted uniformly over the entire domain of the function may be a little too simplified. A more realistic scenario is that the mean, variance, and other aspects of the distribution differ heterogeneously; higher in some regions and lower in others. To study this situation we conduct another set of simulations where the mean and variance are both allowed to vary non-uniformly over the domain, though the range and smoothness are kept constant throughout at $r = 0.4$ and $\nu = 1.0$. More specifically, we generate the mean and standard deviation of $Y$ as two dimensional sine waves centered about 0 and 1, respectively. Then we slowly increase the amplitude of sine waves to make $X$ and $Y$ deviate more in their parameters. The two sine waves are as follows:
\begin{align*}
    \mu(s) &= \left(\frac{\kappa}{2} \sin \left(4\pi s_1 - \pi/2 \right) + 1 \right) \otimes \left(\frac{\kappa}{2} \sin \left(4\pi s_1 - \pi/2 \right) + 1 \right) - 1 \\
    \sigma(s) &=  \left(\frac{\kappa}{2} \sin \left(4\pi s_1 - \pi/2 \right) + 1 \right) \otimes \left(\frac{\kappa}{2} \sin \left(4\pi s_1 - \pi/2 \right) + 1 \right),
\end{align*}
where $s = (s_1, s_2) \in [0, 1] \times [0, 1]$, $\kappa$ is set to vary from 0.05 to 1 in increments of 0.05, and $\otimes$ is the Kronecker product. We fixed the number of replicates to $n = 100$ and $m = 50$ and again used 2000 simulations per $\kappa$ value to estimate the power at $\kappa$.

\begin{figure}
    \begin{center}
    \includegraphics[width=0.8\textwidth,valign=c]{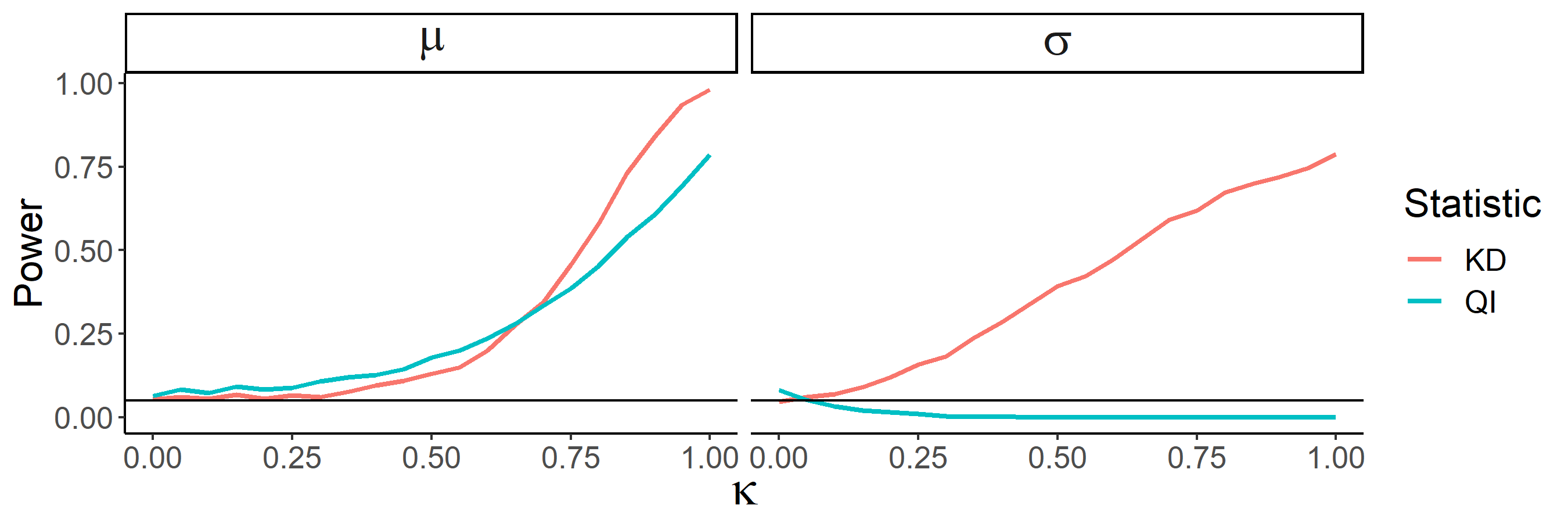}
    \caption{Power functions for $KD$ and $QI$ under heterogeneous differences in the mean and standard deviation between $X$ and $Y$. The parameter $\kappa$ controls the amplitude of the sine waves $\mu(s)$ and $\sigma(s)$, which are centered about 0 and 1 respectively.}
    \label{fig:power_het}
    \end{center}
\end{figure}

Figure \ref{fig:power_het} shows the power functions of $KD$ and $QI$ under heterogeneous mean and standard deviation changes. For detecting mean changes, both $KD$ and $QI$ maintain comparable powers although our test carries more power than the $QI$ test at certain ranges of mean change. It is worth noting that the power curves in this setting appear to be similar to those under the homogeneous mean change which indicates no serious power loss when the mean change is heterogeneous. A huge difference between $KD$ and $QI$ is observed, however, when the standard deviation change is heterogeneous: $KD$ still maintains its power while $QI$ seems to lose power.

Further simulations are again made available in the supplement comparing $KD$, $QI$, $FAD$ and $BAND$ in the heterogeneous case. 
$FAD$ is extremely powerful in detecting mean changes but like $QI$ and $BAND$ it loses all power in detecting heterogeneous variance changes. Our method again shows the highest average power of the four approaches.

\section{Evaluating Proxy Influence in Assimilated CFRs} \label{app}
We now apply the proposed $KD$ statistic to evaluate the influence of proxies on the 2m surface temperature reconstruction by examining the differences between the background and analysis states in PHYDA. In our experiment, the background state consists of a single 100 member ensemble of 2m surface temperature fields that are randomly sampled from a single climate model simulation run. For every year of the reconstruction, the analysis state consists of a 100 member ensemble of 2m surface temperatures (the same 100 randomly drawn ensemble members are selected for both the background and analysis).
We will use our $KD$ statistic to test for distributional differences between the background ensemble and each year's analysis ensemble so as to test for proxy influence during each reconstruction year. Formally, this refers to testing the hypothesis (\ref{eq:test1}) that was formulated in Section \ref{sec:eval}. We will then further subdivide the background and analysis states into 12 regions, corresponding with the five oceans and seven continents, and repeat our analysis on the regions separately. Finally, we investigate how correlation between regions may impact the influence of proxies at the regional level. 



\subsection{Global reconstructions} \label{global}

\begin{figure}[h!]
  \centering
    \includegraphics[width=0.6\textwidth]{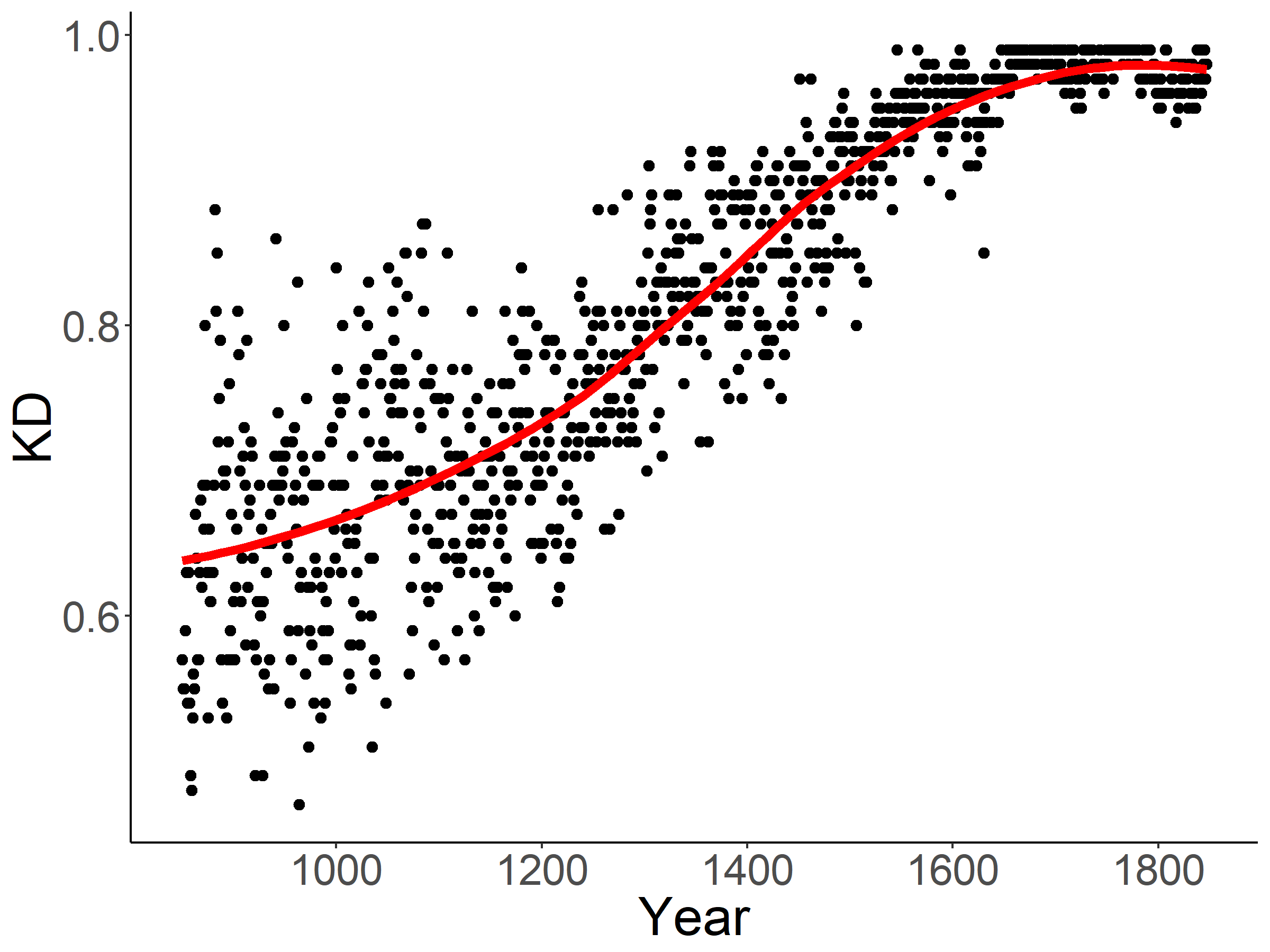}
    \caption{Value of $KD$ over the reconstruction period from 850 CE to 1850 CE. Larger values of $KD$ indicate larger differences between the distribution of the background and analysis states. Red line shows the overall increasing trend of KD. All p-values were less than $6 \times 10^{-10}$ (after FDR adjustment).}
  \label{global_results}
\end{figure}

Figure \ref{global_results} shows the values of $KD$ over time along with their associated p-values. A larger value of $KD$ corresponds to a smaller p-value and more separation between the background and analysis states in their distribution. The p-values were adjusted by the Benjamini-Yekutieli procedure \citep{benjamini2001control} to have a false discovery rate (FDR) of 0.05. The uniformly near zero p-values strongly indicate that the background and analysis are significantly different in distribution each year, which suggests that the proxies indeed change the distribution of the background and have a material influence over the PHYDA reconstructions. Despite the uniformly small p-values, the magnitudes of $KD$ indicate a relatively weak separation between the background and analysis in the beginning followed by a steadily increasing separation over time until the end of the reconstruction period. The apparent rise in separation is caused by the fact that proxy information is sequentially introduced into the reconstruction over time. Over the interval of our analysis (850CE to 1850CE), more proxies become available for assimilation as the reconstruction approaches 1850CE.

\subsection{Regional variation of proxy influence} \label{regional}
\begin{figure}
  \centering
  \subfloat{%
    \includegraphics[width=0.45\textwidth]{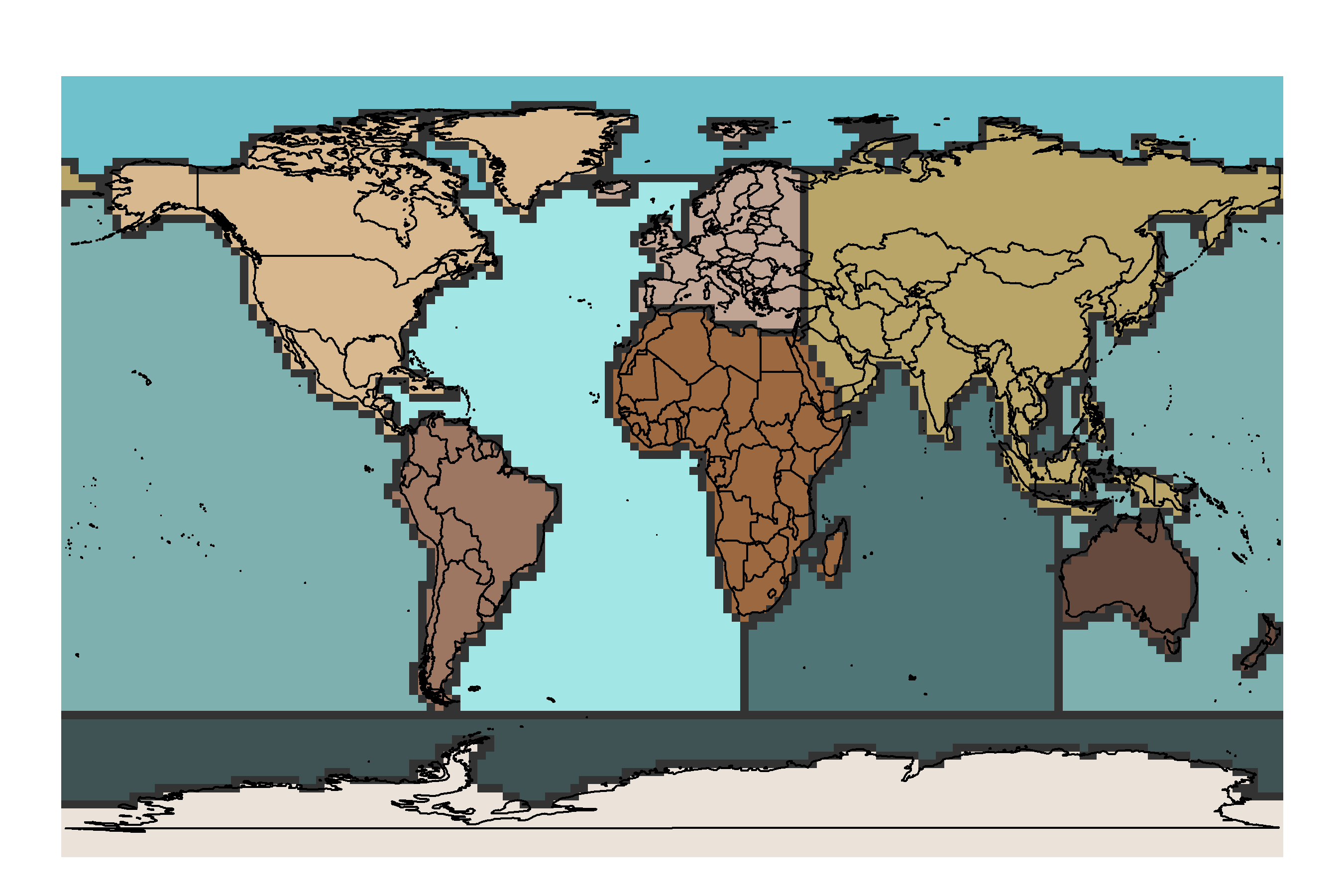}
  }
  \subfloat{%
    \includegraphics[width=0.45\textwidth]{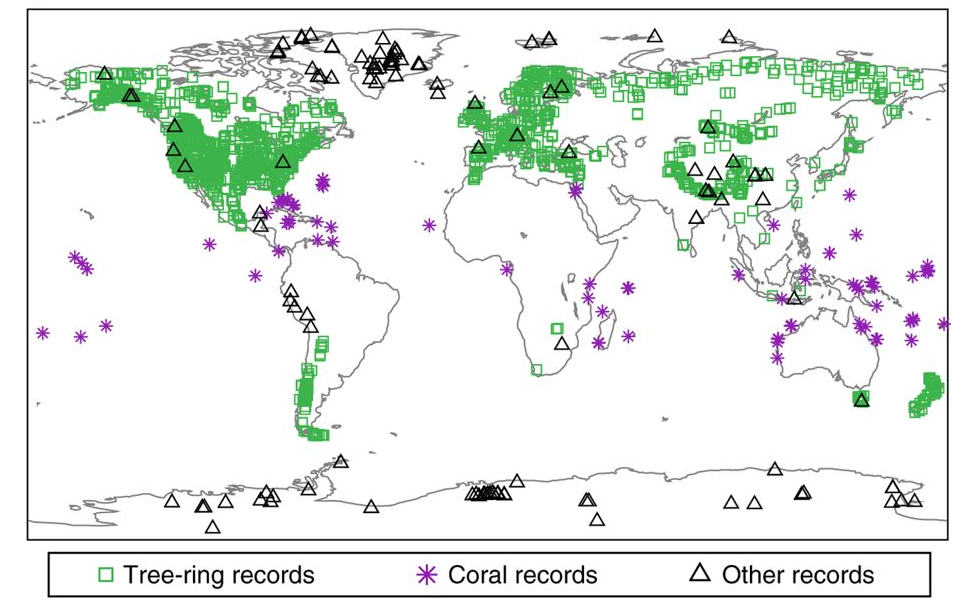}
  }
  \caption{The left panel divides the whole globe into 12 regions marked by different colors and the right panel shows the locations of all of the $n=2978$ proxies used in PHYDA, as in Fig.~1a of \cite{steiger2018PHYDA}. The vast majority of proxies are collected in North America and Europe. Not all displayed proxies are available every year in the reconstruction. More proxies become available as the reconstruction approaches the present day, see Fig.~\ref{fig:temp_dist}.}
  \label{fig:regions}
\end{figure}

Analysis of the global reconstruction is important for establishing the strength of proxy influence at the global level and for confirming the upward trend of the proxy influence. 
A natural next step is to investigate how these effects propagate down to a regional level, namely how proxies impact the temperature reconstruction at the continental and oceanic level.
Proxies are not collected uniformly across all regions as shown in Figure \ref{fig:regions}, so a weaker influence might be expected of the proxies in the poorly sampled regions than those with dense sampling. We therefore use our method to investigate the local influence of proxies.

\begin{figure}
    \begin{center}
    \includegraphics[width=0.80\textwidth,valign=c]{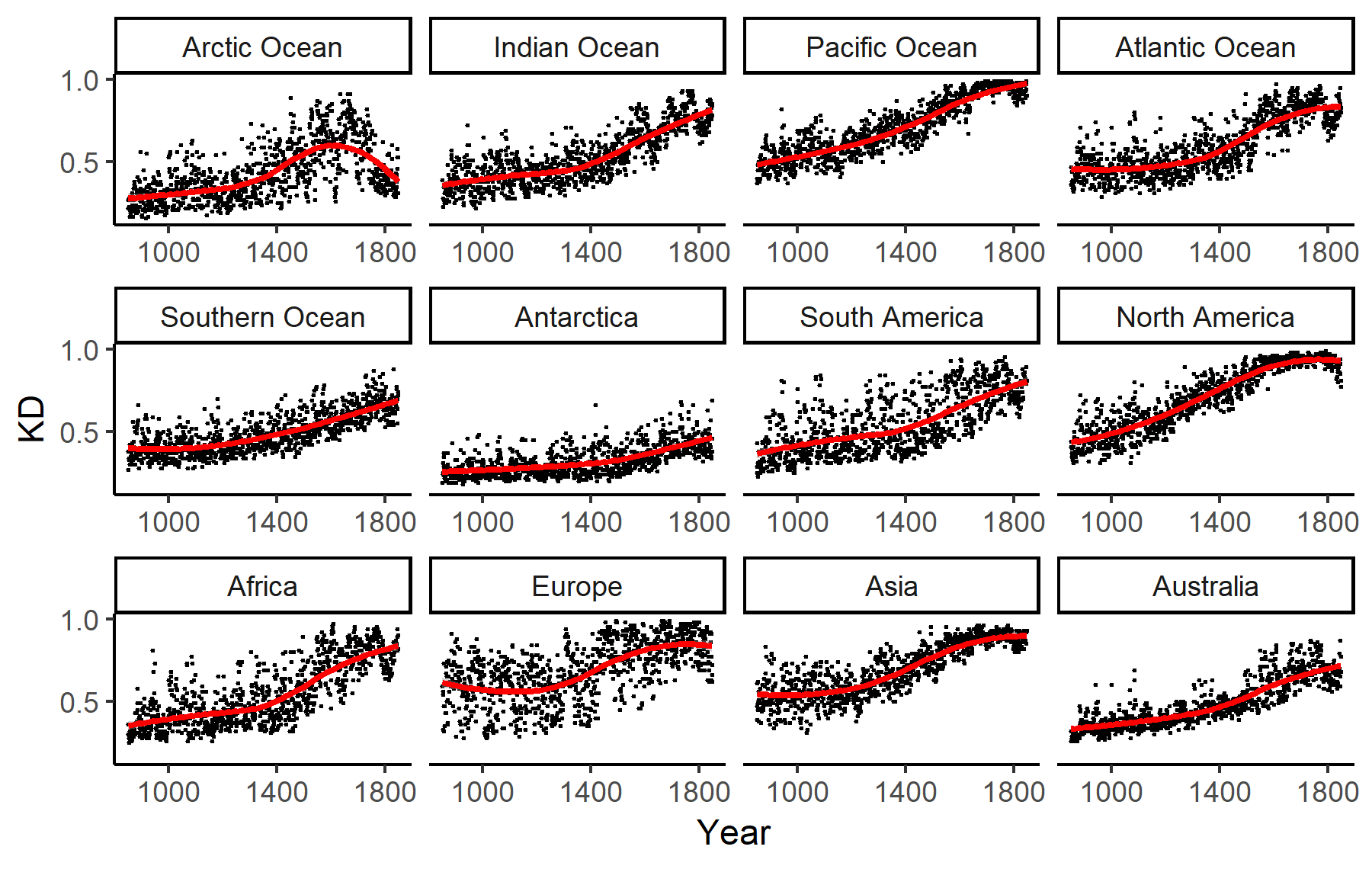}
    \caption{$KD$ over time by region. Regional $KD$ values were computed by measuring differences in the regions of interest within the Global reconstructions. They generally follow the pattern of the global $KD$ values with the exception of the Arctic Ocean. Red lines show the regional trends of KD.}
    \label{fig:effect_regions}
    \end{center}
\end{figure}

We divide the globe into 12 regions corresponding to the five oceans and seven continents as in Figure \ref{fig:regions}. Within each of the twelve regions we apply our test to evaluate the difference between the background and analysis states over the full reconstruction period. 
Analogous to the global study in Section \ref{global}, the progression of $KD$ over time for all regions is summarized in Figure \ref{fig:effect_regions}. It is surprising to see that $KD$ values over all regions share a consistent increasing trend, even for the regions with scarce proxies. Intuitively, we expect that the increasing trend holds only for the regions with abundant proxies because gradually introduced proxy information in those regions will make the analysis states more and more distinct from the background. However, due to the complex dependency structure of the climate system its teleconnections, these regional deficits are likely being mitigated. This result intrigues us to study whether the long range dependence in the background climate states helps to stabilize the reconstruction in data-sparse regions. 

We investigate this conjecture using the correlation maps in Figure \ref{corr}, for which the value at each location describes the strongest correlation, i.e., the maximum $r^2$,  between the temperature time series at this location and every temporally available proxy location during the representative years of 1000, 1400, and 1800 CE, respectively. These maps indicate the maximum potential strength of spatial diffusion of proxy information over time. As more proxies are added towards the present, more global area becomes highly correlated with the proxy locations.

\begin{figure}
    \begin{center}
    \includegraphics[width=0.80\textwidth,valign=c]{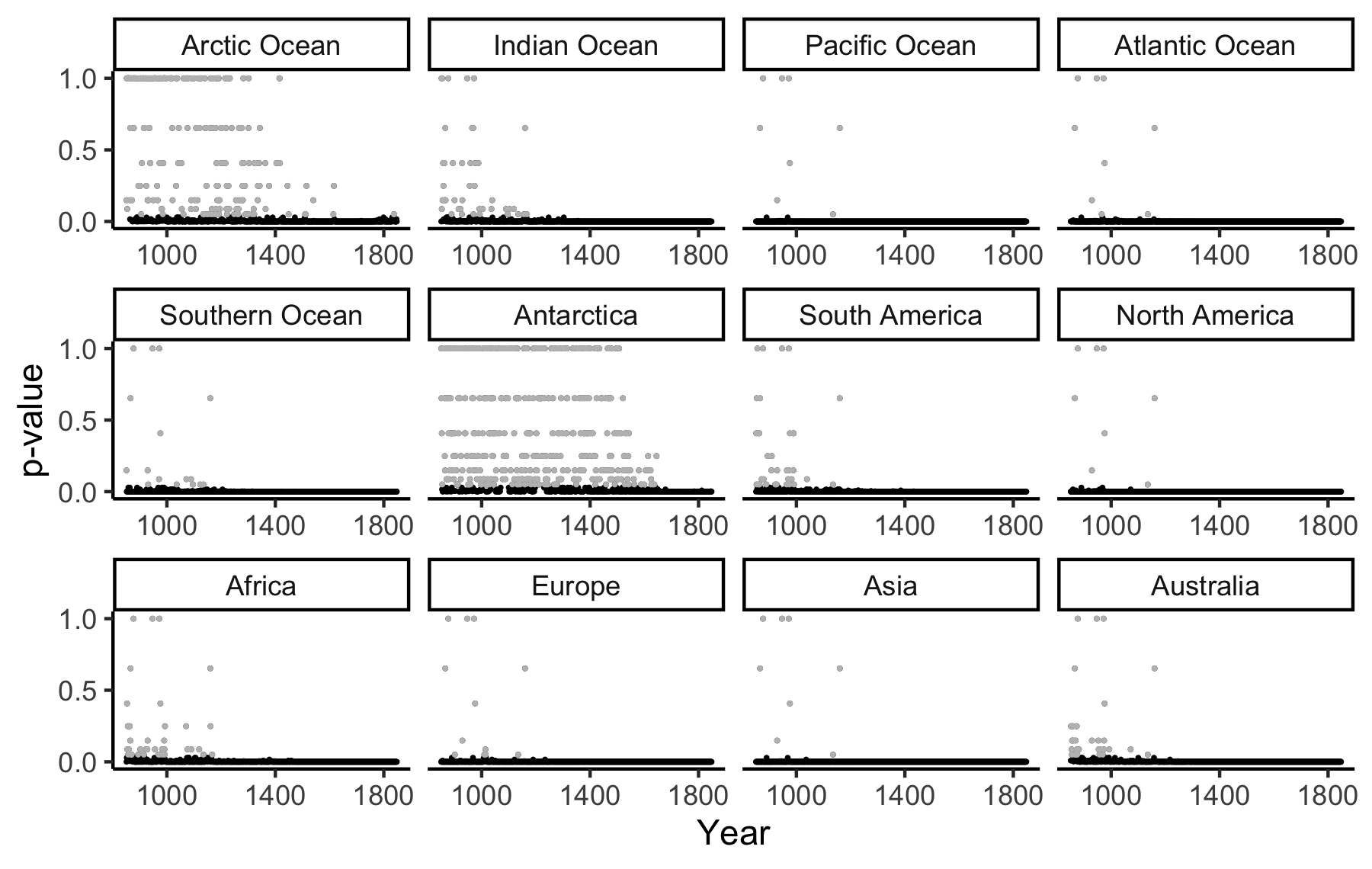}
    \caption{P-values of $KD$ over time by region. Grey points indicate p-values over 0.05 after the Benjamini-Yekutieli FDR adjustment. Except for in the early years, most regions have statistically significant differences between their background and analysis ensembles across the reconstruction. The Arctic Ocean and Antarctica fail to reject in many more cases due to their relatively small size and lack of proxies.}
    \label{fig:pval_regions}
    \end{center}
\end{figure}

\begin{figure}
    \begin{center}
    \includegraphics[width=\textwidth,valign=c]{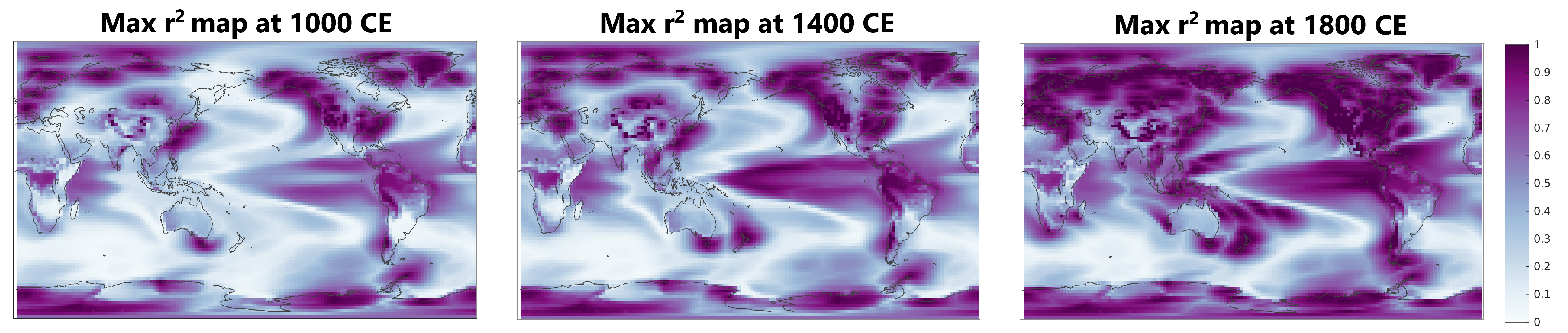}
    \caption{Proxy-point $r^2$ maps for representative years 1000, 1400, and 1800 CE. There is an overall increasing proxy point correlation (purple) in time. This is reflected in the increasing effect sizes seen in regions with little proxy representation.}
    \label{corr}
    \end{center}
\end{figure}

The maps in Figure \ref{corr} are helpful in understanding why some regions such as the Pacific have few proxies but show a high degree of divergence between their background and analysis states. The Pacific is strongly correlated with nearby continental regions such as North America, which has many proxies, due to the El Ni{\~n}o-Southern Oscillation phenomenon. Conversely, Australia, which has few proxies and weak proxy correlations, shows a correspondingly low degree of divergence between its background and analysis states. Regions with densely sampled proxies tend to show high proxy correlation due to proximity with their own proxies, and also a high degree of background-analysis separation, as expected. 

The Arctic and Southern oceans represent two anomalies with regards to their apparent proxy information. Most notably around 1600 CE the Arctic ocean experiences a strong trend reversal in $KD$, just when other regions are experiencing trend increases. This runs counter to the fact that both the number of proxies and the proxy-point correlations shown in Figure \ref{corr} are increasing in the Arctic over this time period. 

\begin{figure}
    \centering
    \includegraphics[width=\textwidth]{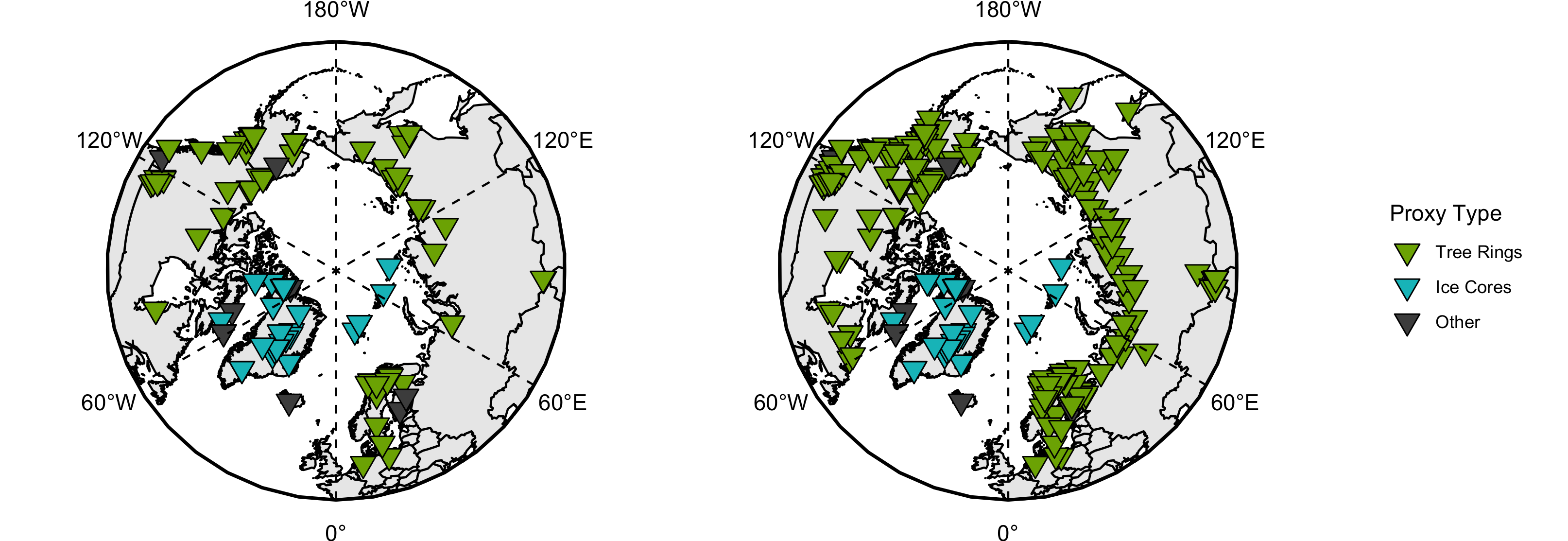}
    \caption{Left: polar plot of proxy sites near the Arctic Ocean (above $50^{\circ}$ latitude) in 1500 CE. Right: proxy sites near the Arctic Ocean in 1700 CE. The number of tree-rings (green) more than triples from 1500 to 1700 CE (75 to 260 sites respectively) without a similar increase in the number of ice cores, marine cores, corals, etc.}
    \label{fig:arctic_tree_rings}
\end{figure}

One possible explanation of this effect is the dramatic increase in the number of Arctic tree-ring records beginning around ~1600 CE, prior to which ice core and sediment records dominate in the Arctic region (see Figure \ref{fig:arctic_tree_rings}).  These latter records are isotope based and have been shown to sample far-field temperature signals across regions of the Arctic Ocean \citep{steiger2017climate}, while several of the isotope records included in the reconstruction are specifically marine based.  It is possible that the isotope records are therefore better samples of the far-field and exclusively marine temperatures that are reconstructed across the Arctic Ocean, relative to the land-based tree-ring records that begin to dominate in the 17th century and sample local temperature conditions.  The inclusion of more widely abundant tree-ring records that are potentially less informative of far-field marine temperatures over the Arctic Ocean may therefore effectively increase the noise in the reconstruction over that region, thus obscuring the contribution from the isotope records and increasing the reliance of the data assimilation on the information from the prior.

Conversely, the Southern ocean has relatively large values of $KD$ when the correlation information in Figure \ref{corr} would lead us to believe that they should be much smaller. The Southern ocean has no local proxies and perhaps has the weakest overall proxy-point correlation strength, yet it experiences a strong and significant divergence between its background and analysis states. This unexpectedly strong divergence may be due to the large amount of moderate correlations observed near the Pacific, along the coast of Antarctica, and off the southern tip of South America. The cumulative effect of those moderate correlations may lead to the results in Figure \ref{fig:effect_regions}.

\section{Discussion and Conclusions} \label{discussion}

Motivated by the newly available PHYDA reconstruction product, we developed a non-parametric statistical test to compare the distributions of the ensembles in the background states and in the analysis states. The PHYDA data product was derived using a DA scheme that merges information from climate model simulations and climate proxies, the latter of which is expected to provide its due influence on the derived analysis fields. However, the nature of the DA approach and the variation of proxy information through time makes it difficult to assess the degree to which the proxies influence the final analysis product.
    
Optimally adding proxy information is one of the principal qualities of a DA-based reconstruction and thus knowing the cumulative effect of adding proxies is of fundamental importance, particularly as DA becomes increasingly popular \citep{franke2017reconstructing,steiger2018PHYDA,tardif2019last}. Before now, testing for significant proxy influence over DA reconstructions has been conducted empirically through validation procedures \citep[e.g.,][]{hakim2016LMR,singh2018insights}. Our test instead provides a direct and powerful way to formally quantify the information added by proxies to the analysis states based on changes in their distribution from the background. By treating each ensemble member in the background and analysis states as continuous two dimensional surfaces, our test statistic based on functional data depth is able to measure the difference in distribution between ensembles in the two states.

Due to the nonparametric nature of functional data depth, our method does not require any distributional or model assumptions on the observations. Our method also does not require that the curves be square integrable, second-order stationary, or even strictly continuous. We showed numerically that the asymptotic distribution of our test statistic converges to or at least is well approximated by the Kolmogorov distribution. Additional simulations in the supplement shows the same conclusion even when the spatial data is from a Non-Gaussian process (see Figures 3 and 4 in the supplement). We also demonstrated that the sizes of our test are well controlled near the nominal level, even under moderate sample sizes, and that our test's powers are highly competitive with the $QI$ test in \citet{quality}. 

Our results provide strong evidence of a clear divergence between the background and analysis states associated with PHYDA. The degree of separation, however, depends greatly on geographical location and time period. An overall upward trend in proxy influence is seen and it is generally maintained even when subdividing the globe into oceanic and continental sub-regions. With the notable exception of the Arctic, these findings are consistent with the fact that proxy information steadily increases as the reconstruction period approaches the present day. This confirms that increasing proxy information is associated with commensurate influence on the assimilated reconstructions, which suggests that the influence of the model prior is minimized as proxy networks become considerably more dense, therefore placing less emphasis on which model should be used to form the prior. This also suggests that more proxies should be collected further back in time to improve reconstruction skill over all parts of the Common Era.

We have also found that, despite the stark imbalance in proxy density in the different geographic regions, most regions exhibit an increasing separation between the background and analysis states. The mitigating effect for the proxy deficit regions is mostly attributable to the long-range dependency structure that proxies and temperatures often display. Some regions such as the Pacific Ocean and South America have very few local proxies but due to their strong overall correlation with other regions they still benefit from proxies collected remotely. These results therefore suggest that the desirable addition of proxy information to data assimilated reconstructions extends beyond the immediate regions where proxies are densely sampled.  This provides credence to the idea that the geographic regions outside of dense proxy sampling may still establish some reconstruction skill, particularly in the last several centuries before the present.  

In addition to the important results for assessing assimilated reconstruction products, our test is much more broadly applicable. Our generic formulation allows it to be applied to any functional data that the depth function can handle, including curves on $\mathbbm{R}$ and higher dimensional functions on $\mathbbm{R}^n$. In our framework, each $X_i$ and $Y_j$ can also be multivariate valued so long as they both map to the same subspace of $\mathbbm{R}^p$. This only changes integration to be over a multivariate depth instead of a univariate depth. Our method can also be useful for comparing image data in medical studies, and meet the increasing demand of comparing simulated climate from different climate models and comparing the simulated climate to observations.

\section*{Acknowledgments}
This research was supported in part by the National Science Foundation through awards OISE-1743738, AGS-1602581, AGS-1602920, and AGS-1805490.
This paper describes objective technical results and analysis. Any subjective views or opinions that might be expressed in the paper do not necessarily represent the views of the U.S. Department of Energy or the United States Government.
This work was supported by the Laboratory Directed Research and Development program at Sandia National Laboratories, a multi-mission laboratory managed and operated by National Technology and Engineering Solutions of Sandia, LLC, a wholly owned subsidiary of Honeywell International, Inc., for the U.S. Department of Energy's National Nuclear Security Administration under contract DE-NA0003525. Lamont contribution no. XXXX.

\section*{Data Availability}
 PHYDA is publicly available at the Zenodo data repository as NetCDF4 files: \url{https://doi.org/10.5281/zenodo.1154913}. The 100 member ensembles of PHYDA used herein are available at: \url{http://clifford.ldeo.columbia.edu/nsteiger/recon_output/phyda_ens/}.

\appendix
\section{Appendix}

\subsection{Proofs}

\textbf{Proposition \ref{asymptotic}}


\begin{proof}
let $P$ be a distribution on $C[0, 1]^p$ and suppose $X = \{X_1,...,X_n\}$ and $Y = \{Y_1,...,Y_n\}$ are two i.i.d samples from $P$. Let $\widehat{F}_n(\cdot)$ and $\widehat{G}_m(\cdot)$ be defined as before with each converging in distribution to $F$, the distribution over $D(\cdot, P)$. Let $x \in X$, then
\begin{align*}
    &\quad \sqrt{\frac{nm}{n+m}} \max_{x \in X}|\widehat{F}_n(x) - \widehat{G}_m(x)| \\
    &\leq \sqrt{\frac{nm}{n+m}} \max_{x \in X}|\widehat{F}_n(x) - F(x)| +  \sqrt{\frac{nm}{n+m}} \max_{x \in X}| F(x) - \widehat{G}_m(x)| \\
    &\simeq \sqrt{m} \max_{x \in X}|\widehat{F}_n(x) - F(x)| +  \sqrt{m} \max_{x \in X}| F(x) - \widehat{G}_m(x)|,
\end{align*}
Because $n \gg m$. By the enforced uniformity of $\widehat{F}_n(x)$ we get that $\max_{x \in X}|\widehat{F}_n(x) - F(x)| = o_p(\frac{1}{\sqrt{n}})$ and so the following upper bound
\begin{align*}
    &\leq o_p(1) +  \sqrt{m} \max_{x \in X}| F(x) - \widehat{G}_m(x)|
\end{align*}
The second term is simply a one sample Kolmogorov-Smirnov statistic so the whole quantity converges to the Kolmogorov distribution.
\end{proof}

\bibliographystyle{chicago}
\bibliography{refs}

\appendix
\section{Appendix}

\subsection{Power comparisons with other tests}

While the quality index is the closest method to ours, it is not the only other method for comparing the distributions of functions. One particularly powerful method is the Functional Anderson-Darling ($FAD$) test of \citep{staicu}. In their paper they demonstrated superior power over all other functional distribution tests except for the Rank based Band Depth Test ($BAND$) of \citep{banddepth}, which was not compared against. We compare our method $KD$ against the Quality Index (QI), $FAD$, and $BAND$ under the same simulation settings as in the main paper. 

\begin{figure}[H]
	\begin{center}
    \includegraphics[width=\textwidth,valign=c]{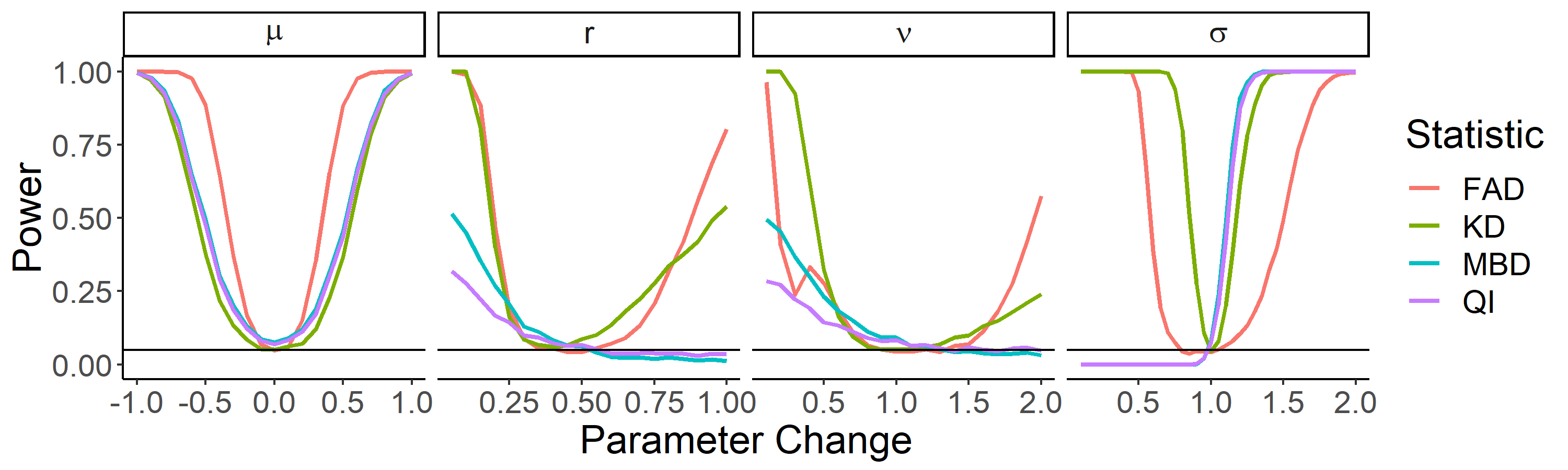}
    \caption{Power of $KD$, $QI$, $FAD$, and $BAND$ in detecting changes in the four parameters in the Gaussian process. Mean, Range and smoothness are presented as shifts of parameters in $Y$ from $X$. Standard deviation is presented as a multiple of standard deviation in $X$.}
    \label{fig:power_const2}
    \end{center}
\end{figure}

\begin{figure}[H]
	\begin{center}
    \includegraphics[width=\textwidth,valign=c]{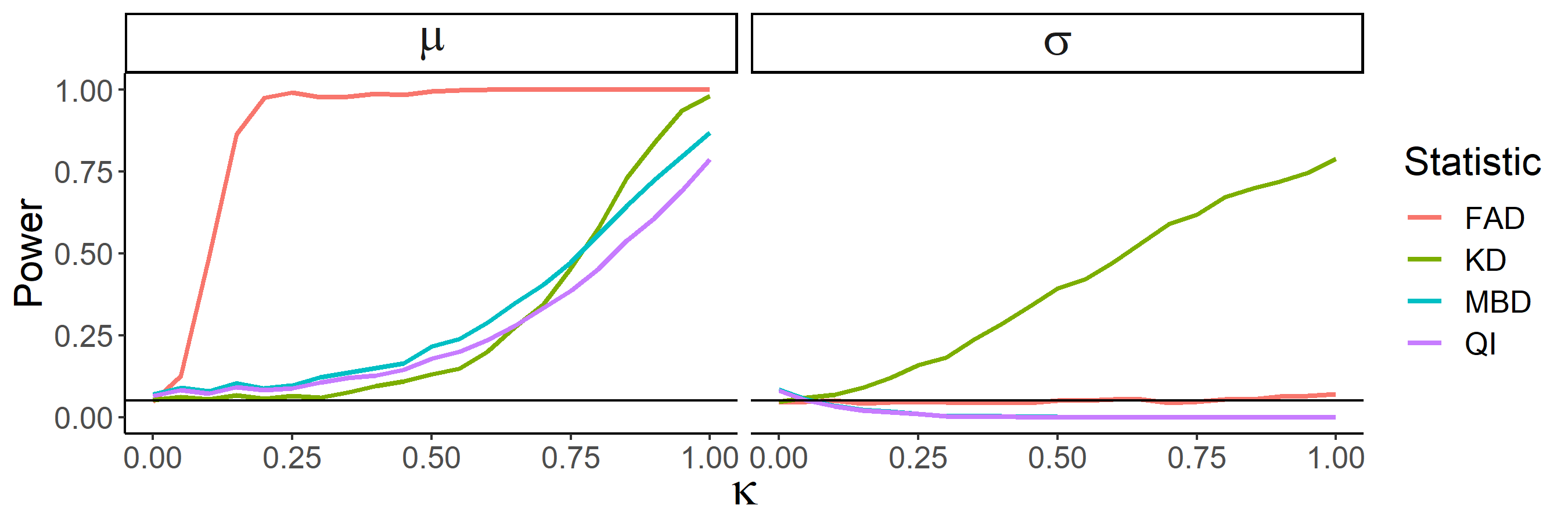}
    \caption{Power of $KD$, $QI$, $FAD$, and BAND in detecting changes in the four parameters in the Gaussian process. Mean, Range and smoothness are presented as shifts of parameters in $Y$ from $X$. Standard deviation is presented as a multiple of standard deviation in $X$.}
    \label{fig:power_het2}
    \end{center}
\end{figure}

The $FAD$ method is extremely powerful against changes in the mean of the data, however compared with the depth based methods its noticeably less powerful against variance changes (Figure \ref{fig:power_const2}). Under the heterogeneous changes (Figure \ref{fig:power_het2}) our test is still the only test to maintain its power in detecting heterogeneous variance changes. 

\subsection{Convergence under a non-Gaussian Process}

Because our test does not depend on any parametric assumptions of the data we wanted to see how convergence, size, and power were maintained when the data came from a markedly Non-Gaussian process. For these simulations we used the same settings as in the main paper's simulations except that the functions were generated with a multivariate t distribution instead of a multivariate Gaussian distribution. We analogously denote these functions as coming from a \textit{t-process}.

\begin{figure}[H]
	\begin{center}
    \includegraphics[width=0.80\textwidth,valign=c]{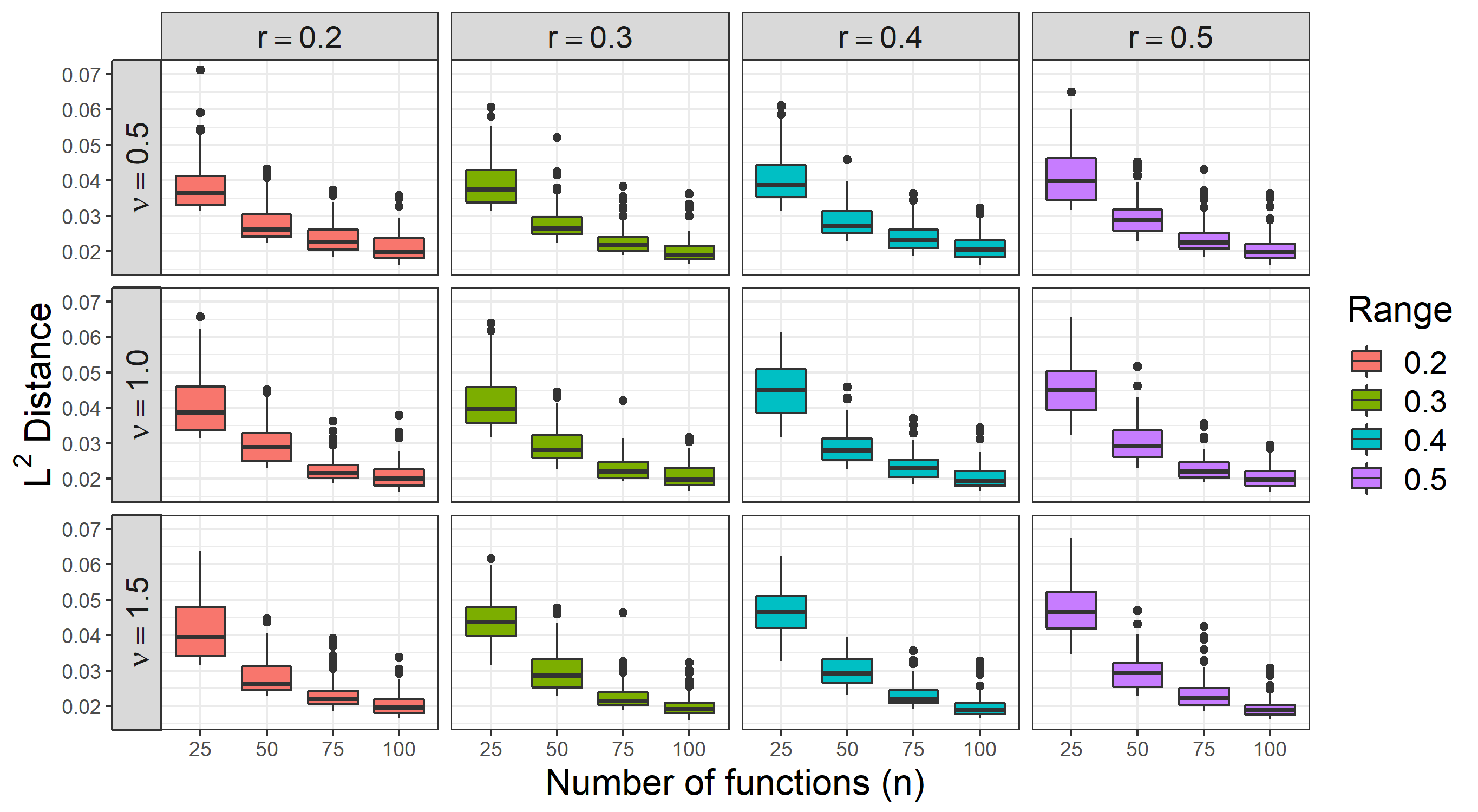}
    \caption{$\Ltwo$ distance between the permutation distribution and the Kolmogorov distribution under 12 different range and smoothness settings. Non Gaussian Process.}
    \label{fig:ngp_l2}
    \end{center}
\end{figure}

\begin{figure}[H]
	\begin{center}
    \includegraphics[width=0.80\textwidth,valign=c]{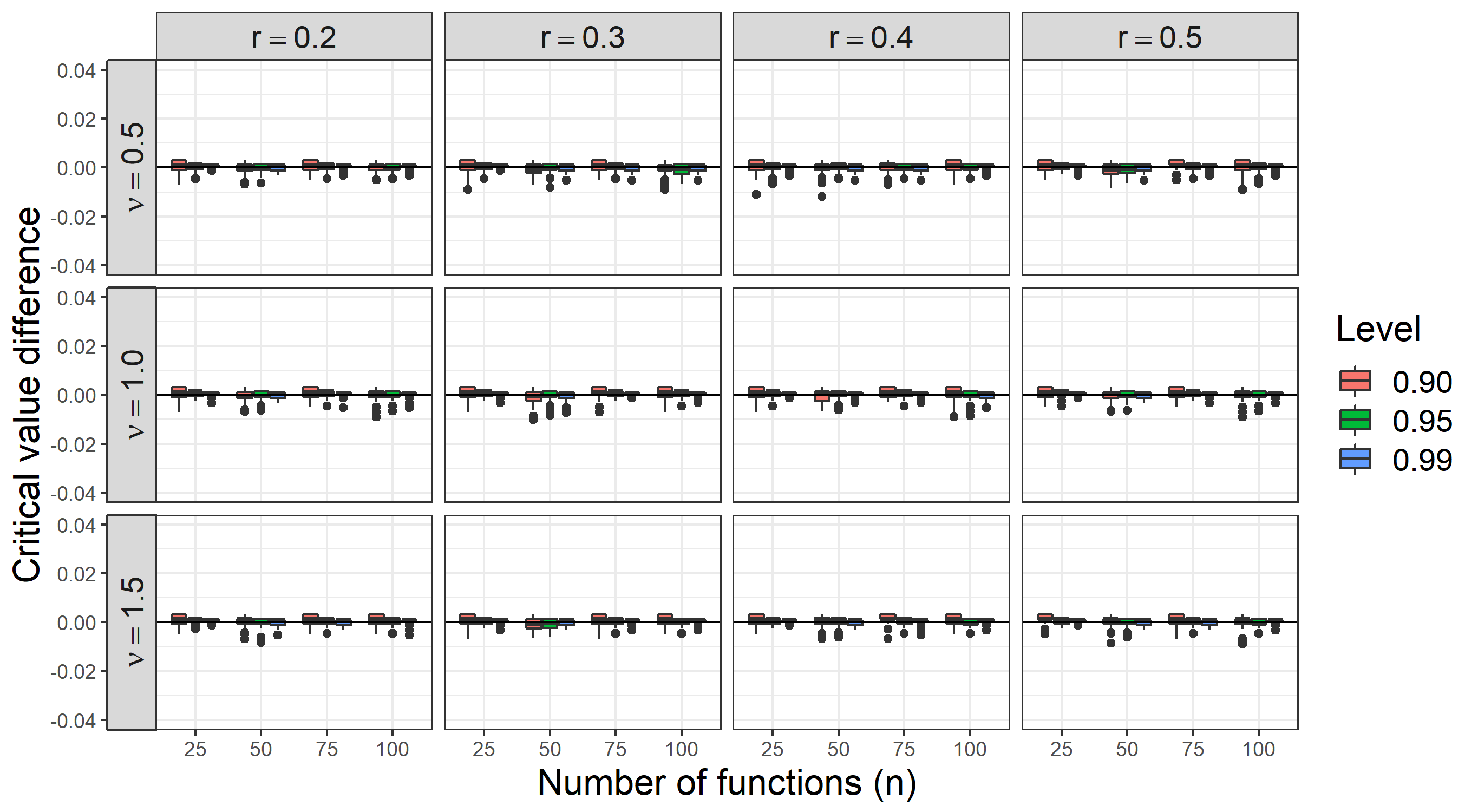}
    \caption{Kolmogorov critical values minus permutation critical values at three common test levels: 0.90, 0.95. 0.99 under 12 different range and smoothness settings. Non Gaussian Process.}
    \label{fig:ngp_crit}
    \end{center}
\end{figure}


Under a t-process, convergence in $\L^2$ is observed to be slower than the corresponding Gaussian process. Critical values, however, are almost immediately unbiased verses their asymptotic counterparts. Together these indicate that the distribution of $KD$ is harder to approximate when the data is heavy tailed, but that this is relatively unimpactful since decisions regarding significance are unaffected by using the asymptotic distribution.

\subsection{Size under a Non-Gaussian Process}

We next looked at the size under t-process data. Size is controlled at relatively the same levels as when Gaussian process data was used. This is due to the critical values of the permutation distribution and the asymptotic distribution being in near agrement, even at small sample sizes. The same pattern of needed sufficient range or smoothness to achieve the nominal level is still observed.

\begin{table}[H]
\scriptsize
\centering
\begin{tabular}{cc|cccc|cccc|cccc}
  \hline
  \multicolumn{2}{c}{} & \multicolumn{4}{|c}{$\nu = 0.5$}& \multicolumn{4}{|c}{$\nu = 1.0$} & \multicolumn{4}{|c}{$\nu = 1.5$}\\ 
 n & m & r = 0.2 & 0.3 & 0.4 & 0.5 & 0.2 & 0.3 & 0.4 & 0.5 & 0.2 & 0.3 & 0.4 & 0.5 \\ 
 \hline
50 & 50 & 0.16 & 0.11 & 0.07 & 0.07 & 0.08 & 0.06 & 0.06 & 0.05 & 0.06 & 0.05 & 0.05 & 0.04 \\ 
&  & (0.25) & (0.19) & (0.14) & (0.14) & (0.14) & (0.12) & (0.11) & (0.10) & (0.12) & (0.11) & (0.10) & (0.08) \\ 
50 & 100 & 0.13 & 0.09 & 0.08 & 0.06 & 0.08 & 0.07 & 0.05 & 0.05 & 0.08 & 0.05 & 0.05 & 0.05 \\ 
&  & (0.29) & (0.19) & (0.18) & (0.14) & (0.18) & (0.14) & (0.11) & (0.09) & (0.15) & (0.10) & (0.10) & (0.09) \\ 
50 & 200 & 0.14 & 0.08 & 0.08 & 0.06 & 0.08 & 0.06 & 0.07 & 0.05 & 0.07 & 0.06 & 0.05 & 0.06 \\ 
&  & (0.34) & (0.22) & (0.19) & (0.16) & (0.18) & (0.12) & (0.12) & (0.10) & (0.15) & (0.11) & (0.11) & (0.10) \\ 
50 & 300 & 0.14 & 0.09 & 0.09 & 0.06 & 0.08 & 0.06 & 0.06 & 0.05 & 0.06 & 0.06 & 0.06 & 0.05 \\ 
&  & (0.36) & (0.26) & (0.22) & (0.17) & (0.20) & (0.16) & (0.13) & (0.12) & (0.17) & (0.12) & (0.10) & (0.10) \\
\hline
100 & 50 & 0.15 & 0.09 & 0.07 & 0.06 & 0.08 & 0.05 & 0.06 & 0.05 & 0.05 & 0.04 & 0.04 & 0.05 \\ 
&  & (0.15) & (0.12) & (0.10) & (0.09) & (0.10) & (0.08) & (0.08) & (0.06) & (0.08) & (0.08) & (0.07) & (0.07) \\ 
100 & 100 & 0.08 & 0.07 & 0.05 & 0.06 & 0.05 & 0.05 & 0.05 & 0.04 & 0.06 & 0.05 & 0.04 & 0.04 \\ 
&  & (0.16) & (0.12) & (0.10) & (0.09) & (0.12) & (0.10) & (0.08) & (0.07) & (0.09) & (0.09) & (0.08) & (0.08) \\ 
100 & 200 & 0.10 & 0.06 & 0.06 & 0.05 & 0.07 & 0.05 & 0.05 & 0.05 & 0.06 & 0.06 & 0.04 & 0.05 \\ 
&  & (0.21) & (0.14) & (0.12) & (0.11) & (0.12) & (0.11) & (0.09) & (0.08) & (0.10) & (0.09) & (0.08) & (0.07) \\ 
100 & 300 & 0.10 & 0.07 & 0.05 & 0.06 & 0.06 & 0.06 & 0.05 & 0.05 & 0.05 & 0.05 & 0.05 & 0.05 \\ 
&  & (0.23) & (0.16) & (0.13) & (0.12) & (0.13) & (0.10) & (0.10) & (0.09) & (0.11) & (0.08) & (0.09) & (0.08) \\
\hline
200 & 50 & 0.15 & 0.10 & 0.08 & 0.07 & 0.08 & 0.07 & 0.05 & 0.06 & 0.07 & 0.06 & 0.05 & 0.06 \\ 
&  & (0.09) & (0.07) & (0.08) & (0.08) & (0.07) & (0.08) & (0.06) & (0.06) & (0.07) & (0.06) & (0.06) & (0.06) \\ 
200 & 100 & 0.10 & 0.08 & 0.06 & 0.05 & 0.06 & 0.05 & 0.05 & 0.04 & 0.05 & 0.05 & 0.05 & 0.05 \\ 
&  & (0.10) & (0.08) & (0.08) & (0.08) & (0.08) & (0.08) & (0.06) & (0.07) & (0.08) & (0.07) & (0.06) & (0.07) \\ 
200 & 200 & 0.08 & 0.06 & 0.06 & 0.05 & 0.06 & 0.05 & 0.05 & 0.04 & 0.06 & 0.06 & 0.05 & 0.05 \\ 
&  & (0.14) & (0.10) & (0.09) & (0.09) & (0.10) & (0.08) & (0.08) & (0.08) & (0.08) & (0.08) & (0.08) & (0.06) \\ 
200 & 300 & 0.09 & 0.07 & 0.06 & 0.07 & 0.06 & 0.06 & 0.06 & 0.05 & 0.05 & 0.05 & 0.05 & 0.06 \\ 
&  & (0.14) & (0.11) & (0.11) & (0.10) & (0.10) & (0.09) & (0.09) & (0.08) & (0.08) & (0.07) & (0.06) & (0.07) \\
\hline
300 & 50 & 0.14 & 0.09 & 0.07 & 0.07 & 0.07 & 0.06 & 0.06 & 0.05 & 0.06 & 0.06 & 0.05 & 0.06 \\ 
&  & (0.08) & (0.07) & (0.06) & (0.07) & (0.06) & (0.06) & (0.05) & (0.05) & (0.07) & (0.05) & (0.06) & (0.06) \\ 
300 & 100 & 0.10 & 0.07 & 0.06 & 0.06 & 0.07 & 0.05 & 0.05 & 0.06 & 0.06 & 0.06 & 0.06 & 0.05 \\ 
&  & (0.09) & (0.08) & (0.06) & (0.07) & (0.08) & (0.07) & (0.06) & (0.06) & (0.06) & (0.06) & (0.06) & (0.06) \\ 
300 & 200 & 0.08 & 0.07 & 0.07 & 0.05 & 0.06 & 0.06 & 0.06 & 0.05 & 0.06 & 0.07 & 0.06 & 0.06 \\ 
&  & (0.11) & (0.09) & (0.08) & (0.07) & (0.08) & (0.07) & (0.06) & (0.07) & (0.07) & (0.08) & (0.06) & (0.06) \\ 
300 & 300 & 0.08 & 0.07 & 0.05 & 0.05 & 0.06 & 0.06 & 0.05 & 0.05 & 0.06 & 0.05 & 0.05 & 0.05 \\ 
&  & (0.10) & (0.09) & (0.08) & (0.08) & (0.08) & (0.08) & (0.07) & (0.07) & (0.06) & (0.07) & (0.05) & (0.07) \\ 
\hline
\end{tabular}
\caption{Size of $KD$ and $QI$ (in parenthesis) under 12 combinations of range, $r$, and smoothness, $\nu$, and 16 combinations of sample sizes, $n$ and $m$, for $X$ and $Y$ respectively. Data was generated from a Non-Gaussian process.}
\label{tab:ngp_size}
\end{table}

\subsection{Power comparisons under a Non-Gaussian Process}

\begin{figure}[H]
	\begin{center}
    \includegraphics[width=\textwidth,valign=c]{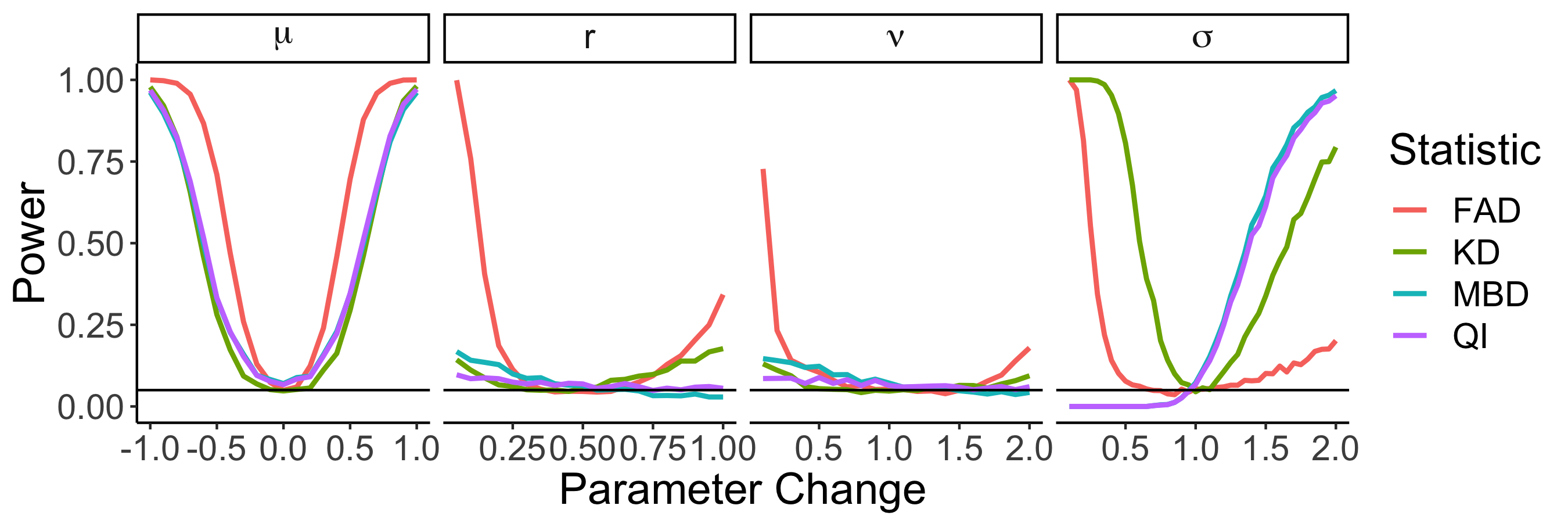}
    \caption{Power of $KD$, $QI$, $FAD$, and $BAND$ in detecting changes in the four parameters in the Gaussian process. Mean, Range and smoothness are presented as shifts of parameters in $Y$ from $X$. Standard deviation is presented as a multiple of standard deviation in $X$.}
    \label{fig:ngp_power_const}
    \end{center}
\end{figure}

\begin{figure}[H]
	\begin{center}
    \includegraphics[width=\textwidth,valign=c]{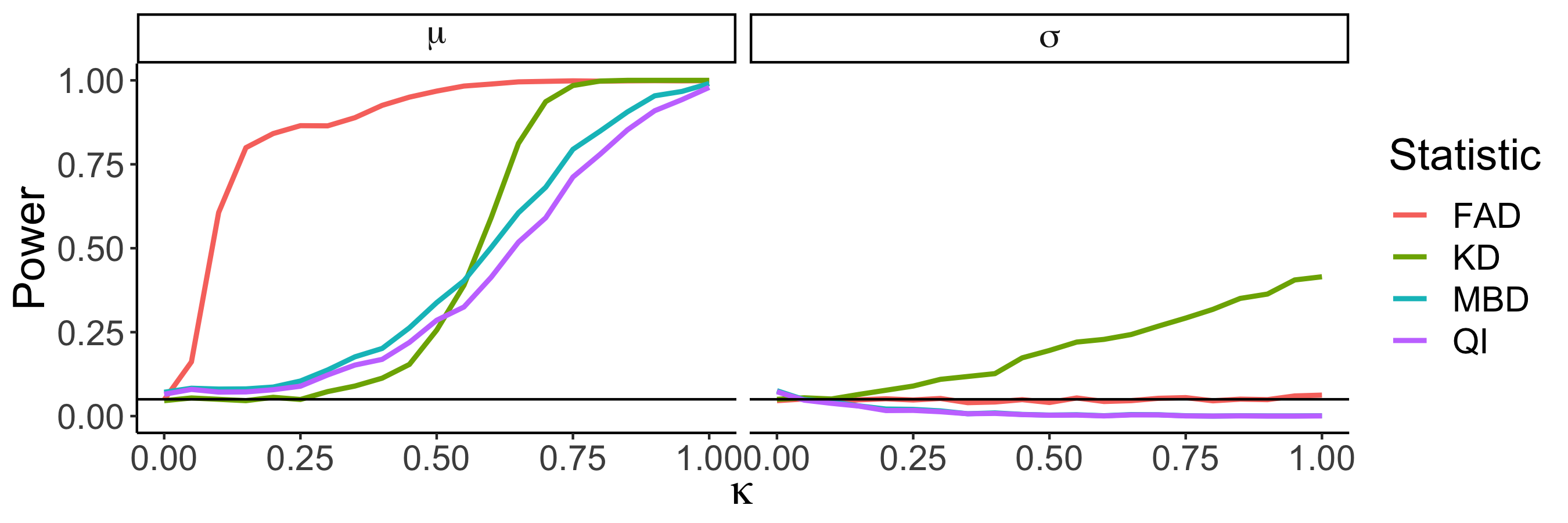}
    \caption{Power of $KD$, $QI$, $FAD$, and $BAND$  in detecting changes in the four parameters in the Gaussian process. Mean, Range and smoothness are presented as shifts of parameters in $Y$ from $X$. Standard deviation is presented as a multiple of standard deviation in $X$.}
    \label{fig:ngp_power_het}
    \end{center}
\end{figure}

Finally we considered power under homogeneous and heterogeneous parameter changes under Non-Gaussian data (t-process). The same settings to test power in the main paper's simulations were against used to generate data. As in the convergence and size simulation the sampled functions were generated from a t-process with 3 degrees of freedom.. The power curves (Figures \ref{fig:ngp_power_const} and \ref{fig:ngp_power_het}) are generally flatter than the corresponding power curves under a Gaussian process, however the relationship between methods remains the same. FAD still dominates detecting changes in the mean and KD, MBD, and QI dominate detecting changes in the standard deviation. All methods lose considerable power in detecting range and smoothness changes. Notably the FAD test ran into computational issues trying to estimate the functional principal components due the t-process frequently generating very outlying curves.

\subsection{FAD v.s. KD on PHYDA}
\begin{figure}
  \centering
  \subfloat{%
    \includegraphics[width=0.45\textwidth]{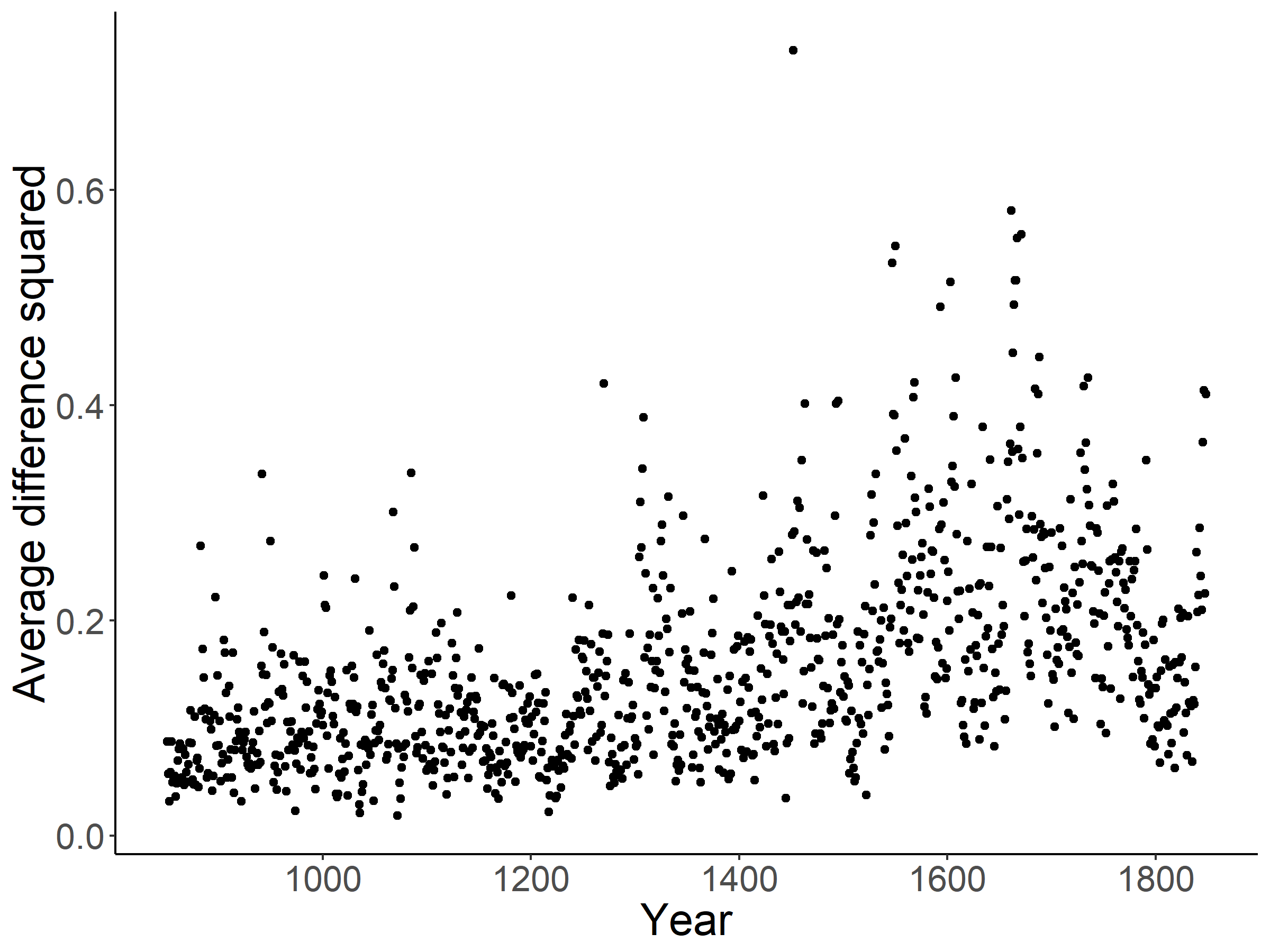}
  }
  \subfloat{%
    \includegraphics[width=0.45\textwidth]{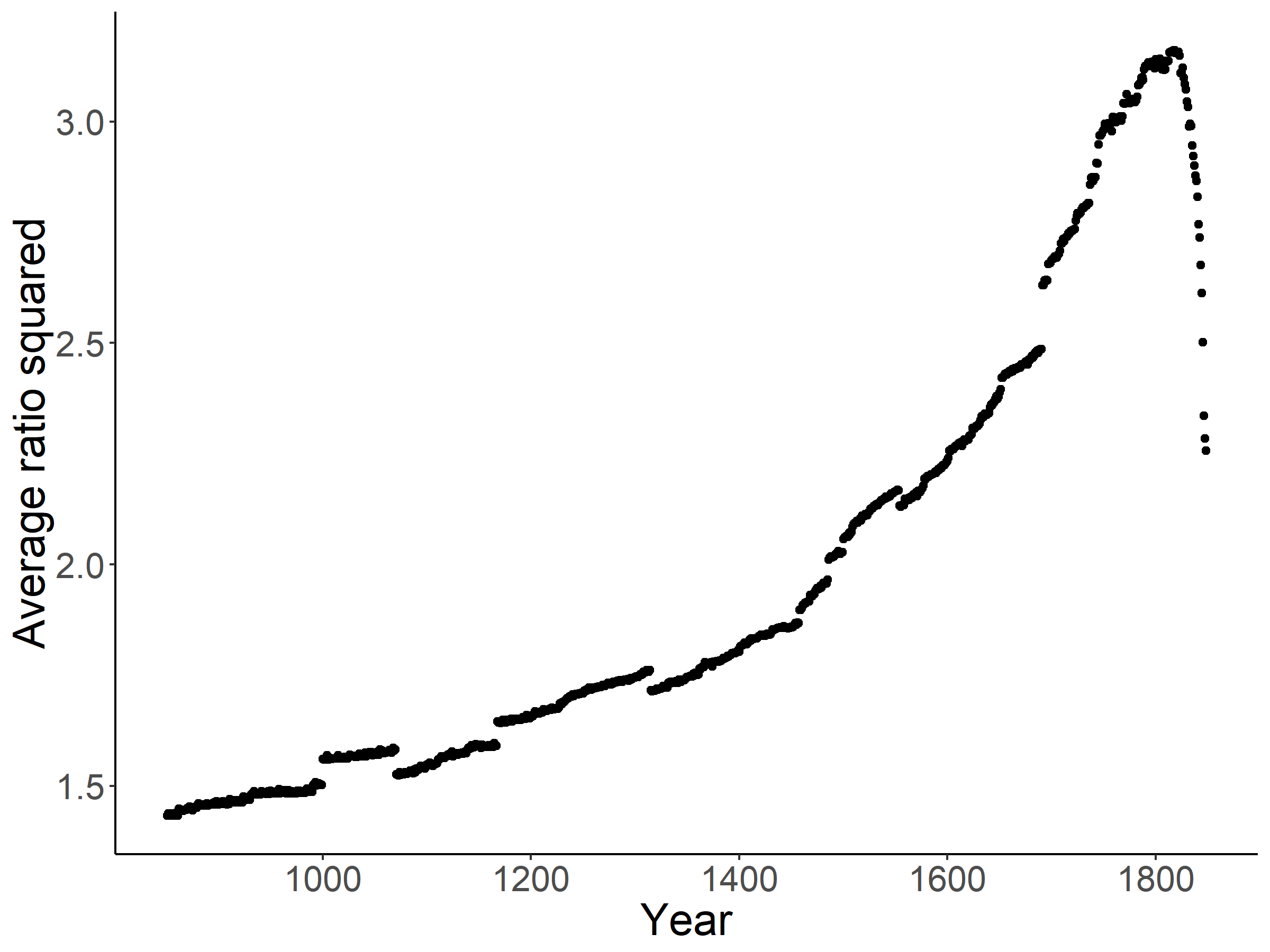}
  }
  \caption{\textbf{Left:} Average squared pointwise mean differences between the background and analysis ensembles for each year in the reconstruction. \textbf{Right:} Average squared pointwise ratio of the background and analysis ensemble standard deviations for each year in the reconstruction.}
  \label{fig:mean_var}
\end{figure}

The preceding power plots show that their is no clear dominating method, between $FAD$ and $KD$ across all of the parameters in the Gaussian and Non-Gaussian simulations. $FAD$ clearly detects mean differences better and $KD$ clearly detects standard deviation differences better. This is particularly true in the case of heterogeneous mean and variance changes under a t-process (Figure \ref{fig:ngp_power_het}), i.e. the more realistic setting. We argue that because $FAD$ fails to detect heterogeneous changes in the variance, it misses out on the crucial finding in our data analysis, namely that the analysis ensembles become more distinct from the background over time. These changes appear to be primarily driven by a downward trend in the variance of the analysis state (see Figure \ref{fig:mean_var}).  

\begin{figure}
  \centering
  \subfloat{%
    \includegraphics[width=0.45\textwidth]{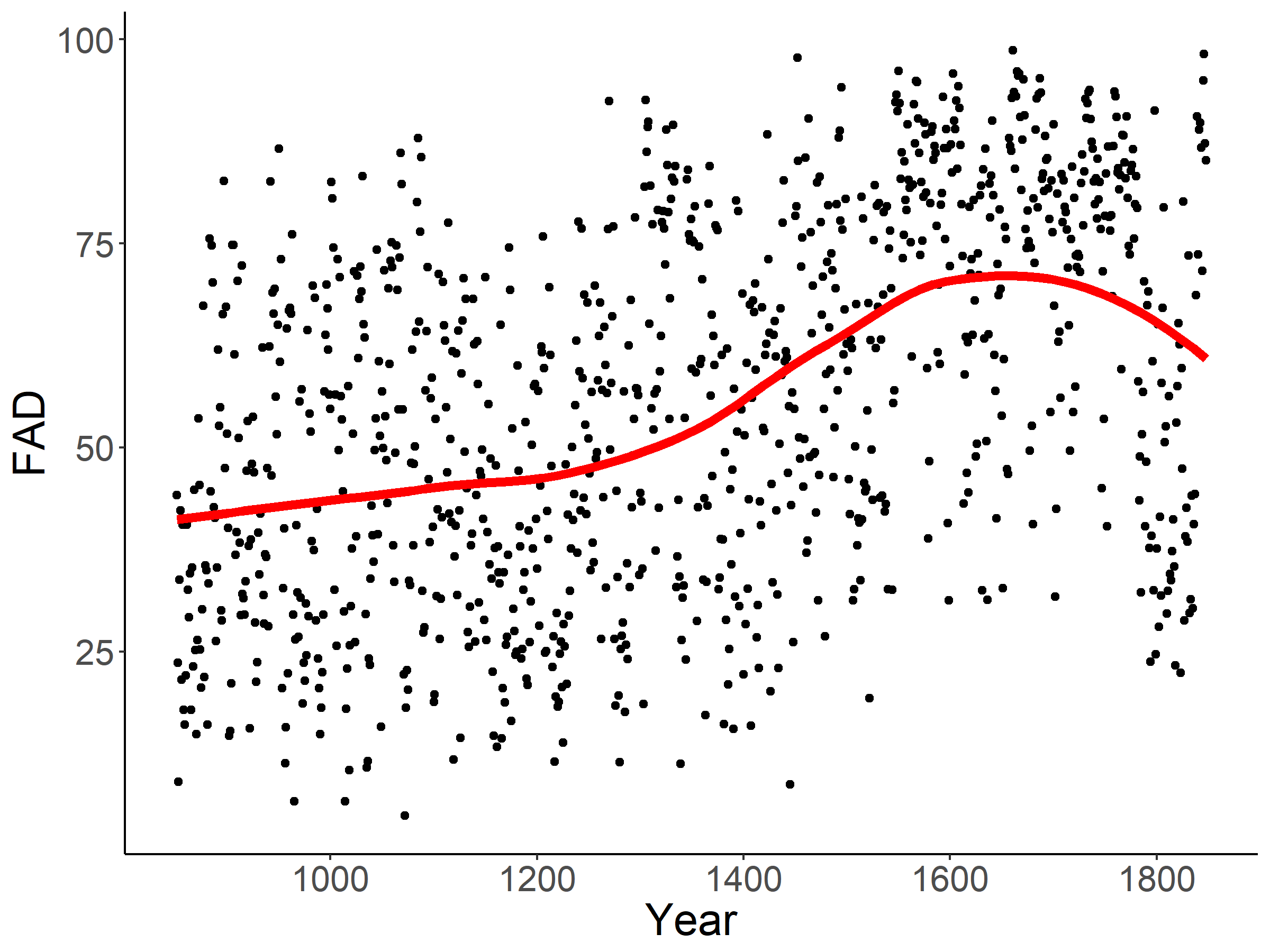}
  }
  \subfloat{%
    \includegraphics[width=0.45\textwidth]{results/effect_over_time.png}
  }
  \caption{$FAD$ vs $KD$ values on the PHYDA climate data over the reconstruction period 850CE to 1850CE. Both tests detect significant distribution changes, but $FAD$ is primarily driven by the mean differences. $KD$ derives its value from the mean changes, the increase standard deviation changes, and higher moment changes not displayed here.}
  \label{fig:effects}
\end{figure}

As can be seen in Figure \ref{fig:mean_var}, the average difference between the background and analysis remains relatively constant over time. Because the averages differences are even slightly different from 0, $FAD$ has no issue with detecting a significant difference. The real differentiator is how the ratio of the variances changes over time. With the exception of the very end of the reconstruction, the average variance ratio increases almost monotonically. This pattern reveals that one of the primary effects of including additional proxies is a reduction in uncertainty. This near monotonic increase in uncertainty reduction is largely reflected in the associated time series of K values (Figure \ref{fig:effects}). If we compare against the values of $FAD$ over time (Figure \ref{fig:effects}) we can see that it does not register this aspect of the distribution change. $FAD$ generally only follows the trend of the mean differences, while $KD$ follows both.
\end{document}